\documentstyle[12pt,epsf]{article}

\newcommand{\beq}{\begin{equation}}
\newcommand{\eeq}{\end{equation}}
\newcommand{\bea}{\begin{eqnarray}}
\newcommand{\eea}{\end{eqnarray}}
\newcommand{\rar}{\rightarrow}

\newcommand{\lan}{\langle}
\newcommand{\ran}{\rangle}

\newcommand{\bk}{{\bf k}}
\newcommand{\bN}{{\bf N}}
\newcommand{\bY}{{\bf Y}}
\newcommand{\bone}{{\bf 1}}
\newcommand{\balpha}{\mbox{\boldmath$\alpha$}}
\newcommand{\bsigma}{\mbox{\boldmath$\sigma$}}
\newcommand{\cL}{{\cal L}}

\newcommand{\cH}{{\cal H}}
\newcommand{\bcN}{\mbox{\boldmath$\cal N$}}
\newcommand{\stheta}{S_{\theta}}
\newcommand{\ctheta}{C_{\theta}}
\newcommand{\sphi}{S_{\phi}}
\newcommand{\cphi}{C_{\phi}}

\begin{document}

\font\fortssbx=cmssbx10 scaled \magstep2
\hbox to \hsize{
\includegraphics{uwlogo.ps}
\hskip.5in \raise.1in\hbox{\fortssbx University of Wisconsin - Madison}
\hfill$\vcenter{\hbox{\bf MADPH-95-880}
            \hbox{April 1995}}$ }
\vskip 2cm
\begin{center}
\Large
{\bf On the instantaneous Bethe-Salpeter equation} \\
\vskip 0.5cm
\large
 M. G. Olsson and  Sini\v{s}a Veseli \\
\vskip 0.1cm
{\small \em Department of Physics, University of Wisconsin, Madison,
	\rm WI 53706} \\
\vspace*{+0.2cm}
 Ken Williams \\
{\small \em Continuous Electron Beam Accelerator Facility \\
	Newport News, VA 29606, USA \\
	and \\
\vspace*{-0.2cm}
Physics Department, Hampton University, Hampton, VA 29668}
\end{center}
\thispagestyle{empty}
\vskip 0.7cm

\begin{abstract}
We present a systematic algebraic and numerical investigation
of the instantaneous Bethe-Salpeter equation. Emphasis
is placed on confining interaction kernels of the
Lorentz scalar, time component vector, and full  vector
types. We explore stability of the solutions and Regge
behavior for each of these interactions, and conclude that only time
component vector confinement leads to normal Regge structure and stable
solutions.
\end{abstract}

\newpage

\section{Introduction}
The Bethe-Salpeter equation \cite{bib:bethe} follows from the general
principles
of quantum field theory \cite{bib:gellmann}. The instantaneous
form of the Bethe-Salpeter equation, known as the Salpeter equation
\cite{bib:salpeter}, avoids difficulties related to the
relative time degree of freedom, and is believed to provide a
firm framework for the discussion of bound state problems.
Almost all knowledge of
the solutions to the Salpeter equation is restricted to that
sector where at least one of the
constituent masses is large, in which case use of the
so called reduced Salpeter equation is justified.
Relatively little work has been done on the algebraic
properties \cite{bib:smith,bib:murota,bib:yao} of the full Salpeter
equation, and even less on its numerical solution. A few years
ago Laga$\ddot{\rm e}$ \cite{bib:lagae1} proposed
a formalism which shows considerable promise for the systematic
investigation of the full Salpeter equation. He also examined
several confinement models \cite{bib:lagae2}, and concluded that
the confining potential is not a scalar on the basis of
non-linear Regge behavior of the equal mass solutions.
More recently, M$\ddot{\rm u}$nz et. al. \cite{bib:munz}
investigated a linear Lorentz
scalar or alternatively a time component vector confining kernel combined
with an effective interaction proposed by 't Hooft
from instanton effects in QCD \cite{bib:thooft}. They found that
there was no convincing parametrization for the confining
kernel that leads to linear Regge trajectories
and also yields  spin orbit terms of the correct sign.
They have also concluded that a scalar
confining kernel does not lead to stationary solutions
for higher angular momenta or small constituent
masses. This conclusion inspired
Parramore and Piekarewicz to perform a stability analysis of the
variational solutions to the Salpeter equation in the
pseudoscalar channel \cite{bib:parramore}.  They concluded
 that time component vector confinement is stable
with respect to the increase in number of basis states,
but they found the
existence of imaginary eigenvalues for scalar confinement.
One may argue though \cite{bib:lagae3} that these
authors have not considered the norm of the solutions in their analysis.

In this paper we extend Laga$\ddot{\rm e}$'s method, correcting
a small but important
algebraic error and exploring for the first time
the full vector interaction kernel. We consider the nature of the
solutions to the full Salpeter equation for interaction kernels of the
time component Lorentz vector, scalar, and full vector types.
The variational (Galerkin) method is used to investigate the
reality of eigenvalues and stability of the
solutions with the above Lorentz interactions.
We  have taken the norm of the solutions
into account in our analysis, since only states with positive norm have
direct physical significance.
We have also investigated the Regge behavior for the
above mentioned kernels, and extended the analysis
done in \cite{bib:lagae2} to  heavy-light systems.
We also find that only the time component
vector interaction leads to stable variational solutions
and has normal Regge behavior.

This paper is organized as follows. In Section \ref{sec:rev}
we review some general properties of the Salpeter equation
and the reduction to a system of coupled radial equations. The numerical
results are discussed for each of the Lorentz kernels in
Section \ref{sec:res}, and our conclusions are summarized in
Section \ref{sec:con}. Appendix \ref{app:eqs} contains
the complete coupled radial equations for the kernels considered in
this paper, as well as discussion of important limiting cases.
The techniques of our numerical solution are
 covered in more detail in Appendix \ref{app:ns}.

\section{Reduction to radial equations}
\label{sec:rev}

In this section we briefly review Laga$\ddot{\rm e}$'s
elegant formalism \cite{bib:lagae1}
for the reduction
of the full Salpeter equation to a system of equations involving only
radial wave functions. Our purpose here is to establish notation,
and to correct a small inconsistency in the
Laga$\ddot{\rm e}$'s derivation
which, for example, leads to a couple of sign errors in the
radial equations (5.7) of \cite{bib:lagae1}. The correction
does not  affect the equal mass case, which was
the subject of Laga$\ddot{\rm e}$'s numerical examples \cite{bib:lagae2}.

We start from the Salpeter equation for a fermion-antifermion system
in the CM frame of the bound state,
\beq
M\chi = H_{1}\chi - \chi H_{2} + \int \frac{d^{3}\bk'}{(2\pi)^{3}} V(\bk-\bk')
(\Lambda_{+}^{1}\Gamma_{1}\chi'\Gamma_{2}\Lambda_{-}^{2}
-\Lambda_{-}^{1}\Gamma_{1}\chi'\Gamma_{2}\Lambda_{+}^{2})\ ,\label{eq:salpeq}
\eeq
with notation $f=f(\bk),\ f'=f(\bk')$. In the above
 equation $V(\bk-\bk')$
is a scalar function with  Fourier transform $V(r)$ in the case
of a Lorentz vector kernel, and $-V(r)$ in the case of a Lorentz
scalar kernel.
The Salpeter amplitude $\chi$ describing a mesonic bound state
$|B\ran $ is defined by
\beq
\chi(\bk) = \lan 0|\psi_{1}(\bk)\psi_{2}^{\dagger}(\bk)|B\ran\ ,
\eeq
$\Gamma_{1}$ and $\Gamma_{2}$ are  $4\times 4$ matrices
corresponding to an interaction kernel with Lorentz structure
$\gamma^{0}\Gamma_{1}\otimes \gamma^{0}\Gamma_{2}$, while
the $H_{i}$'s and $\Lambda_{\pm}^{i}$'s are generalized Dirac Hamiltonians
and energy projection operators, given by
\bea
H_{i}(\bk) &=& A_{i}(\bk)\balpha\cdot \hat{\bk} + B_{i}(\bk)\beta\ ,\\
\Lambda_{\pm}^{i} &=&\frac{E_{i}(\bk)\pm H_{i}(\bk)}{2E_{i}(\bk)}\ ,
\eea
with $E_{i}(\bk) = \sqrt{A_{i}(\bk)^{2}+B_{i}(\bk)^{2}}$. In this paper
we restrict ourselves to  constituent quarks of masses $m_{i}$,
so that
\bea
A_{i}(\bk) &=&k\ ,\\
B_{i}(\bk) &=&m_{i}\ ,\\
E_{i}(\bk)&=&\sqrt{m_{i}^{2}+\bk^{2}}\ .
\eea

Using properties of projection operators it can be easily shown
that the Salpeter amplitude satisfies the constraint condition
\beq
\frac{H_{1}}{E_{1}}\chi + \chi \frac{H_{2}}{E_{2}}=0\ .\label{eq:con}
\eeq
Taking this into account,
the norm of the Salpeter amplitude \cite{bib:smith,bib:yao}
can be written as
\beq
||\chi ||^{2}=\int \frac{d^{3}\bk '}{(2\pi)^{3}}
{\rm Tr}\left[\chi^{\dagger}\frac{H_{1}}{E_{1}}\chi\right]\ ,\label{eq:norm}
\eeq
and is related to the normalization of bound states as
\beq
||\chi||^{2}=\frac{1}{(2\pi)^{3}} \lan B| B\ran\ .
\eeq
Using (\ref{eq:salpeq}) inside of (\ref{eq:norm}) one obtains
\beq
M||\chi||^{2} = \int \frac{d^{3}\bk}{(2\pi)^{3}}[E_{1}+E_{2}]\
{\rm Tr}[\chi^{\dagger}\chi] +
\int \frac{d^{3}\bk}{(2\pi)^{3}} \int \frac{d^{3}\bk'}{(2\pi)^{3}}
V(\bk -\bk')\ {\rm Tr}[\chi^{\dagger}\Gamma_{1}\chi'\Gamma_{2}]\ .
\label{eq:var}
\eeq
This equation will be used for obtaining radial equations from the
variational principle as outlined in \cite{bib:lagae1}.

Now, if we expand the Salpeter amplitude as
\beq
\chi=\cL_{0} + \cL_{i}\rho_{i} + \bcN_{0}\cdot \bsigma +
\bcN_{i}\cdot \rho_{i}\bsigma\ ,\label{eq:ampl}
\eeq
using 16 Hermitian matrices whose squares are unity ($1,\rho_{i},
\bsigma,\rho_{i}\bsigma$) as defined in
\cite{bib:yao}, it is then easily seen that
the constraint (\ref{eq:con})
can be satisfied by expressing the 16 components of $\chi$ ($\cL$'s and
$\bcN$'s) in terms of eight functions ($L_{1},L_{2},\bN_{1},\bN_{2}$)
in the following way:
\bea
\cL_{0}&=& \stheta(\hat{\bk}\cdot \bN_{2})\ ,\nonumber\\
\cL_{1}&=& \sphi L_{1}\ ,\nonumber\\
\cL_{2}&=& i\ctheta L_{2}\ ,\nonumber\\
\cL_{3}&=& -\cphi (\hat{\bk}\cdot \bN_{1})\ ,\nonumber\\
\bcN_{0}&=& \stheta L_{2} \hat{\bk} + i\cphi(\hat{\bk}\times \bN_{2})\ ,
\label{eq:con1}\\
\bcN_{1}&=& \sphi  \hat{\bk} (\hat{\bk} \cdot \bN_{1}) -
 \ctheta\hat{\bk}\times(\hat{\bk}\times \bN_{1})\ ,\nonumber\\
\bcN_{2}&=& i[\ctheta  \hat{\bk} (\hat{\bk} \cdot \bN_{2}) -
 \sphi\hat{\bk}\times(\hat{\bk}\times \bN_{2})]\ ,\nonumber\\
\bcN_{3}&=& -\cphi L_{1} \hat{\bk} - i\stheta(\hat{\bk}\times \bN_{1})\ .
\nonumber
\eea
Here we have used notation
\bea
\sphi = \sin{\phi}\ ,\ \cphi=\cos{\phi}\ ,\\
\stheta = \sin{\theta}\ ,\ \ctheta=\cos{\theta}\ ,
\eea
with angles $\phi$ and $\theta$ defined as
\beq
\phi=\frac{\phi_{1}+\phi_{2}}{2}\ ,\
\theta=\frac{\phi_{2}-\phi_{1}}{2}\ ,
\eeq
while the $\phi_{i}$'s are defined through
\beq
\cos{\phi_{i}}=\frac{A_{i}}{E_{i}}\ ,\
\sin{\phi_{i}}=\frac{B_{i}}{E_{i}}\ .
\eeq

At this point we have departed from Laga$\ddot{\rm e}$ in one small
detail. Namely, we have redefined the function $L_{1}$ from
equation (4.10) in \cite{bib:lagae1}, so that
$L_{1}^{our}=-L_{1}^{Laga\ddot{e}}$. The reason for
doing so is
that now, using (\ref{eq:ampl}) and
(\ref{eq:con1}) in the expression for the
norm (\ref{eq:norm}), one can obtain
\beq
||\chi||^{2}=4\int\frac{d^{3}\bk}{(2\pi)^{3}}[
L_{2}^{*}(\bk)L_{1}(\bk) + L_{1}^{*}(\bk)L_{2}(\bk)
+\bN_{2}^{*}(\bk)\cdot \bN_{1}(\bk) +
\bN_{1}^{*}(\bk)\cdot \bN_{2}(\bk)]\ ,\label{eq:norm2}
\eeq
which is equation (4.13) from \cite{bib:lagae1}.
Using Laga$\ddot{\rm e}$'s definition of $L_{1}$ would lead
to  minus signs in front of terms $L_{2}^{*}(\bk)L_{1}(\bk)$ and
$L_{1}^{*}(\bk)L_{2}(\bk)$ in (\ref{eq:norm2}). This small
inconsistency of equations (4.10) and (4.13) from \cite{bib:lagae1}
leads to some incorrect signs in the final form of the radial
equations for states with parity $P=(-1)^{J+1}$.
In the equal mass case the terms with incorrect signs
vanish, so that the numerical results obtained in \cite{bib:lagae2}
are not affected.

Now we proceed to obtain the radial equations. We first express $L_{i}$
and $\bN_{i}$ in terms of spherical harmonics and vector
spherical harmonics (for an extensive discussion of
generic wave functions
and the identification of the quantum
numbers of the bound states the reader
is again referred to \cite{bib:lagae1}), so that
\bea
L_{i}(\bk) &=& L_{i}(k)Y_{JM}(\hat{\bk})\ ,\\
\bN_{i}(\bk) &=& N_{i-}(k)\bY_{-}(\hat{\bk}) +
N_{i0}(k)\bY_{0}(\hat{\bk}) +
N_{i+}(k)\bY_{+}(\hat{\bk})\ ,
\eea
where $\bY_{-}$, $\bY_{0}$, and $\bY_{+}$, stand for
$\bY_{JJ-1M}$, $\bY_{JJM}$, and $\bY_{JJ+1M}$, respectively.
We also introduce the functions $n_{i+}$ and $n_{i-}$, defined as
\beq
\left[ \begin{array}{c}
       n_{i+} \\
      n_{i-}
	\end{array} \right]
= \left[\begin{array}{cc}
\mu & \nu \\
    -\nu & \mu  \end{array}\right]
\left[\begin{array}{c}
	N_{i+} \\
      N_{i-} \end{array}\right] ,
\eeq
with
\bea
\mu &=& \sqrt{\frac{J}{2J+1}}\ ,\ \\
\nu &=& \sqrt{\frac{J+1}{2J+1}}\ .
\eea
Using these definitions, together with properties of spherical
and vector spherical harmonics, we find from (\ref{eq:con1})
\bea
\cL_{0}&=& \stheta n_{2-}Y_{JM}\ ,\nonumber\\
\cL_{1}&=& \sphi L_{1}Y_{JM}\ ,\nonumber\\
\cL_{2}&=& i\ctheta L_{2}Y_{JM}\ ,\nonumber\\
\cL_{3}&=& -\cphi n_{1-}Y_{JM}\ ,\nonumber\\
\bcN_{0}&=& \cphi n_{2+}\bY_{0}
+ (\nu \cphi N_{20}+\mu \stheta L_{2})\bY_{-}
+ (\mu \cphi N_{20} - \nu \stheta L_{2})\bY_{+}\ ,
\label{eq:con2}\\
\bcN_{1}&=& \ctheta N_{10}\bY_{0}
+ (\mu \sphi n_{1-} + \nu \ctheta n_{1+})\bY_{-}
+ (-\nu \sphi n_{1-} + \mu \ctheta n_{1+})\bY_{+}\ ,\nonumber\\
\bcN_{2}&=& i[\sphi N_{20}\bY_{0}
+ (\nu \sphi n_{2+} + \mu \ctheta n_{2-})\bY_{-}
+ (\mu \sphi n_{2+} - \nu \ctheta n_{2-})\bY_{+}]\ ,\nonumber\\
\bcN_{3}&=& -\stheta n_{1+}\bY_{0}
+ (-\mu\cphi L_{1} -\nu\stheta N_{10})\bY_{-}
+ (\nu\cphi L_{1} -\mu\stheta N_{10})\bY_{+} \ .
\nonumber
\eea
In the above formulas everything is
expressed in terms of radial functions (e.g. $L_{i}=L_{i}(k)$, and so on).
Let us briefly review the quantum numbers of the states
that these radial functions
represent (parity $P$, charge conjugation $C$, and $^{2S+1}L_{J}$):
\beq
\begin{array}{llll}
L_{1},L_{2} & P=(-1)^{J+1} & C=(-1)^{J} & ^{1}J_{J} \\
N_{10},N_{20} & P=(-1)^{J+1} & C=(-1)^{J+1} & ^{3}J_{J} \\
n_{1+},n_{2+},n_{1-},n_{2-}& P=(-1)^{J} & C=(-1)^{J} & ^{3}(J\pm 1)_{J}
\end{array}
\eeq

Substituting (\ref{eq:con2}) in the expression for the norm
(\ref{eq:norm2}), and using
the angular integrals summarized in \cite{bib:lagae1}, we
find
\bea
||\chi||^{2}&=&4\int_{0}^{\infty} \frac{k^{2}dk}{(2\pi)^{3}}
[L_{1}^{*}L_{2}+ L_{2}^{*}L_{1}
+ N_{10}^{*}N_{20} + N_{20}^{*}N_{10} \nonumber \\
&+&n_{1+}^{*}n_{2+} + n_{2+}^{*}n_{1+}
+ n_{1-}^{*}n_{2-} + n_{2-}^{*}n_{1-}
]\ .\label{eq:varnorm}
\eea
Similarly, for the kinetic energy part of (\ref{eq:var})
we get
\bea
&&\hspace*{-8mm}\int \frac{d^{3}\bk}{(2\pi)^{3}}[E_{1}+E_{2}]\ {\rm Tr}
[\chi^{\dagger}(\bk)\chi(\bk)] =
4 \int_{0}^{\infty} \frac{k^{2}dk}{(2\pi)^{3}}[E_{1} + E_{2}]
\ \times \hspace{+4.4cm}
\nonumber \\
&&\hspace*{-2mm}[L_{1}^{*}L_{1}+ L_{2}^{*}L_{2}
+ N_{10}^{*}N_{10} + N_{20}^{*}N_{20}
+ n_{1+}^{*}n_{1+} + n_{2+}^{*}n_{2+}
+ n_{1-}^{*}n_{1-} + n_{2-}^{*}n_{2-}]\ .\label{eq:varke}
\eea
Each type of kernel must be treated separately. Here, for
example, we give kernel part
of (\ref{eq:var}) for the interaction of the
form $\gamma^{0}\otimes\gamma^{0}$ ($\Gamma_{1}=\Gamma_{2}=1$):
\bea
&&\hspace{-1cm}\int \frac{d^{3}\bk}{(2\pi)^{3}}
\int \frac{d^{3}\bk'}{(2\pi)^{3}}
V(\bk -\bk')\ {\rm Tr}[\chi^{\dagger}\Gamma_{1}\chi'\Gamma_{2}]=
4 (2\pi) \int_{0}^{\infty} \frac{k^{2}dk}{(2\pi)^{3}}
\int_{0}^{\infty} \frac{k'^{2}dk'}{(2\pi)^{3}}
\Bigl\lbrace \hspace{+1.3cm}
\nonumber \\
\hspace{-8mm}&&\hspace{-3mm}L_{1}^{*}[
\sphi V_{J} \sphi'L_{1}' +
\cphi (\mu^{2} V_{J-1} +\nu^{2}V_{J+1})\cphi'L_{1}'
+\mu\nu \cphi(V_{J-1}-V_{J+1})\stheta'N_{10}']\hspace{1cm}
\nonumber \\
\hspace{-8mm}&+&\hspace{-3mm} L_{2}^{*}[
\ctheta V_{J} \ctheta'L_{2}' +
\stheta (\mu^{2} V_{J-1} +\nu^{2}V_{J+1})\stheta'L_{2}'
+\mu\nu \stheta (V_{J-1}-V_{J+1})\cphi'N_{20}']
 \nonumber \\
\hspace{-8mm}&+&\hspace{-3mm} N_{10}^{*}[
\ctheta V_{J} \ctheta'N_{10}' +
\stheta (\nu^{2} V_{J-1} +\mu^{2}V_{J+1})\stheta'N_{10}'
+\mu\nu \stheta(V_{J-1}-V_{J+1})\cphi'L_{1}']
\nonumber \\
\hspace{-8mm}&+&\hspace{-3mm} N_{20}^{*}[
\sphi V_{J} \sphi'N_{20}' +
\cphi (\nu^{2} V_{J-1} +\mu^{2}V_{J+1})\cphi' N_{20}'
+\mu\nu \cphi(V_{J-1}-V_{J+1})\stheta'L_{2}']
 \label{eq:varpe}\\
\hspace{-8mm}&+&\hspace{-3mm} n_{1+}^{*}[
\stheta V_{J} \stheta'n_{1+}' +
\ctheta (\nu^{2} V_{J-1} +\mu^{2}V_{J+1})\ctheta'n_{1+}'
+\mu\nu \ctheta(V_{J-1}-V_{J+1})\sphi'n_{1-}']
\nonumber \\
\hspace{-8mm}&+&\hspace{-3mm} n_{2+}^{*}[
\cphi V_{J} \cphi'n_{2+}' +
\sphi (\nu^{2} V_{J-1} +\mu^{2}V_{J+1})\sphi'n_{2+}'
+\mu\nu \sphi (V_{J-1}-V_{J+1})\ctheta'n_{2-}']
\nonumber \\
\hspace{-8mm}&+&\hspace{-3mm} n_{1-}^{*}[
\cphi V_{J} \cphi'n_{1-}' +
\sphi (\mu^{2} V_{J-1} +\nu^{2}V_{J+1})\sphi'n_{1-}'
+\mu\nu \sphi (V_{J-1}-V_{J+1})\ctheta'n_{1+}']
\nonumber \\
\hspace{-8mm}&+&\hspace{-3mm} n_{2-}^{*}[
\stheta V_{J} \stheta'n_{2-}' +
\ctheta (\mu^{2} V_{J-1} +\nu^{2}V_{J+1})\ctheta'n_{2-}'
+\mu\nu \ctheta (V_{J-1}-V_{J+1})\sphi'n_{2+}']\ \Bigr \rbrace
\nonumber\ ,
\eea
with $V_{L}$  defined as
\beq
V_{L}(k,k') = 8\pi \int_{0}^{\infty} r^{2}dr V(r) j_{L}(kr)j_{L}(k'r)\ .
\label{eq:vdef}
\eeq

At this point
we can obtain the radial equations by taking variations
of (\ref{eq:varnorm}), (\ref{eq:varke}) and (\ref{eq:varpe}),
with respect to $L_{1}^{*}(k),L_{2}^{*}(k)\ldots,n_{2-}^{*}(k)$. The
resulting equations for the
$\gamma^{0}\otimes\gamma^{0}$ kernel,
as well as for the $\bone\otimes \bone$ and
$\gamma^{\mu}\otimes\gamma_{\mu}$ kernels, are summarized in
Appendix \ref{app:eqs}.

Of course, one can obtain these equations also by straightforward
substitution of (\ref{eq:ampl}) and (\ref{eq:con2}) into the Salpeter
equation (\ref{eq:salpeq}), and then taking the trace
after  multiplication of the
resulting equation with the appropriate matrices.
 The angular integrals which one needs
can be worked out easily using the definition of vector
spherical harmonics, the expression for $Y_{l_{1}}^{m_{1}}(\Omega)
Y_{l_{2}}^{m_{2}}(\Omega)$, and the general properties of the Clebsch-Gordan
coefficients, as given in the Appendix C of \cite{bib:messiah}. However,
Laga$\ddot{\rm e}$'s  method reviewed in this section is much more
simple and elegant.

\section{Numerical results}
\label{sec:res}

As outlined in Appendix \ref{app:ns}, one solves the radial equations
by expanding the wave  functions in terms of a complete set of
basis states, which depend on a variational parameter $\beta$.
This expansion is then truncated  to a finite
number of basis states. In this way, a set of coupled radial
equations can be transformed into a matrix equation,
$\cH \psi = M\psi$. The eigenvalues $M$ of the matrix $\cH$ will depend
on $\beta$, and by looking for the extrema of $M(\beta)$, one
can find the bound states. If the calculation is stable,
increasing the number of basis states used will decrease
the  dependence of the eigenvalues
on $\beta$. The regions of $\beta$ with
the same eigenvalues should thus enlarge.

A stability analysis of the variational solutions
for the pseudoscalar states has been recently performed
in \cite{bib:parramore}. Using the fact that the Salpeter equation
can be cast in a  form identical in structure to a
random-phase-approximation (RPA)  equation, the
authors of \cite{bib:parramore}
have employed the same formalism developed by Thouless in his study
of nuclear collective excitations \cite{bib:thouless}, to perform
a stability analysis of the Salpeter equation with Lorentz
time component vector and scalar confining kernels. They find the presence
of instability, manifested by the appearance of imaginary
eigenvalues, in the case of scalar confinement. On the other
hand,  they find no
such evidence in the case of time component vector confinement.

Since matrix $\cH$ is not symmetric, its eigenvalues are not
 guaranteed to be real. However, as noted in
\cite{bib:lagae1,bib:munz2}, the reality of eigenvalues
follows from the reality of the norm and of
the right hand side of (\ref{eq:var}) when $\Gamma_{1}$ and $\Gamma_{2}$
are hermitian, unless the norm is zero. Physically
acceptable solutions must have positive (and non-zero) norm.
One may argue \cite{bib:lagae3}
that the authors of \cite{bib:parramore} have not
taken this into account. Therefore, we find it worthwhile to
 examine the
stability of the variational solutions to the
Salpeter equation by taking the norm into account, i.e.
by rejecting states with negative or zero norm.

In the following we consider the above issues for
three different kernels,
$\gamma^{0}\otimes\gamma^{0}$,
$\gamma^{\mu}\otimes\gamma_{\mu}$, and
$\bone\otimes \bone$.

\subsection{$\gamma^{0}\otimes\gamma^{0}$ kernel [time component
Lorentz vector]}

The case of the $\gamma^{0}\otimes\gamma^{0}$ confining
kernel ($V(r)=ar$) was found
\cite{bib:parramore}
to be stable with respect to an increase in the number of the
basis states, even with very small quark masses. In order
to verify this, we have performed a similar calculation for the
pseudoscalar case $J^{PC}=0^{-+}$, using $a = 0.2\ GeV^{2}$
and zero mass quarks, with as many as 50 basis states.
Results for the lowest three physical states
(with positive norm) are shown in figure \ref{fig:bv0}. One can see that,
as number of  basis states increases, plateaus with the
same eigenvalues enlarge and there is no sign of instability.
The same calculation was performed with states
of higher angular momentum (even as high as $J=20$),
and again results are the same. Therefore, we confirm the
conclusion reached by the authors of \cite{bib:parramore}
that time component vector confinement is well behaved and suitable
for a variational solution. As an additional check of
our programs we have reobtained all the numerical values
for the time component vector confinement with several different
quark masses, which are given in
 Table 1 of  \cite{bib:parramore}.

In Figures \ref{fig:rv0} and \ref{fig:r2v0} we show
the leading and the first few daughter Regge trajectories
in the case of $P=-C$ mesons in the light-light and heavy-light
systems. As expected, we obtained slopes of
$\frac{1}{8a}$ for the light-light, and
$\frac{1}{4a}$ for the heavy-light systems.

In  Figure \ref{fig:ff} we plot radial wave functions
for the lowest lying $S$ and
$P$ waves in coordinate space. We have used
$V(r)=a r$ ($a=0.2\ GeV^{2}$) and
 zero mass quarks. The wave functions were normalized
so that $||\chi||^{2}=\frac{(2\pi)^{3}}{4}$.

Finally, it is a well known fact that the Salpeter equation
does not reduce to the Dirac equation in the limit where one
quark mass becomes infinite \cite{bib:cl}. What is needed is
an interaction which allows the existence of single pair terms.
These terms arise from kernels involving crossed ladder
diagrams. Therefore, it is interesting to compare
the exact solution of the Dirac equation with the
time component vector Coulomb potential ($V(r)=-\frac{\kappa}{r}$)
with the solution of
the Salpeter equation in the heavy-light limit.
In the Figure \ref{fig:dv0} we plot the
Coulomb energy for these two equations as a
function of the light quark mass, and for  three different
values of $\kappa$. For example, for the light quark mass
of $m_{1}=0.3\ GeV$ and for $\kappa = 0.75$, the effect is
of the order of magnitude of $10\ MeV$. We note that
for small $m_{1}$ this system becomes spatially large
and weakly bound.

\subsection{$\bone\otimes \bone$ kernel [Lorentz scalar]}

As far as stability of the scalar confinement is concerned,
things are completely different. The authors of
\cite{bib:parramore} claim that imaginary
eigenvalues occur as they increase number of basis states,
even for  large values of
the quark mass. For the particular choice of $m_{1}=m_{2}=0.9\ GeV$ and
$a=0.29 \ GeV^{2}$ they find that increasing the number
of basis states from 20 to 25 leads to the first occurrence of
imaginary eigenvalues (for the pseudoscalar
state). Using these parameters we find
no imaginary eigenvalues, even with as many as 50 basis states. However,
as shown in Figure \ref{fig:bs1} scalar confinement
in the full Salpeter equation does have a stability problem. As soon
as basis states having a large enough momentum components
are included into calculation  instabilities occur.
As seen in Figure \ref{fig:bs1}, if one includes only 25 basis
states, the three lowest states with positive norm have
all well defined plateaus in variational parameter $\beta$. But
as soon as we go
 from 25 to 35 basis the third state with positive norm
develops instability. Of course,
for the ground state this instability
occurs later. In order to see the magnitude of this effect
we have enlarged the scale for the behavior
of the ground state plateau in the Figure \ref{fig:bs2}.  As one can
see, increasing the number of  of basis states from $39$ to $40$
leads to a  decrease of energy in one
small region of $\beta$. Naturally, these problems
occur much earlier with smaller quark masses.
To gain some insight into the nature of this instability, we show
in Figure \ref{fig:ff1}  the behavior of the
wave functions as
the instability occurs. We have taken the same parameters
as for the previous two figures, and chosen $\beta=1.538\ GeV$
(in the middle of the instability from the Figure \ref{fig:bs2}).
As one might expect,
with $39$ basis states the radial wave functions still preserve
behavior characteristic of an  $S$-wave, which is lost
for the solutions
with $40$ basis states.
It is also interesting
to note that $\beta$ near the edge of the unstable
region (Figure \ref{fig:bs2}) yields wave functions
appear to be identical to the ones in the middle of it.

As far as
imaginary eigenvalues obtained from
the Salpeter equation with scalar confinement are concerned, we have indeed
found these
 with very small quark masses. However,
states with such eigenvalues always have zero norm, and have to be
rejected. Therefore, in this case we do not agree with
 \cite{bib:parramore}, and support conclusions reached
by \cite{bib:lagae1,bib:munz2} that physical states
will have positive norm and positive energies. However, we
do agree with  \cite{bib:parramore} that the Salpeter
equation with scalar confinement does have a stability problem.
Let us also briefly mention that if one squares the ground state
energy from Figure \ref{fig:bs2}, one obtains $6.750\ GeV^{2}$, a
number to which the calculation of   \cite{bib:parramore}
converges before instability occurs (Table 2 in \cite{bib:parramore}).
We do not want to speculate on the reasons why
calculations done in  \cite{bib:parramore} are much less stable
than ours (for example, with parameters
used for Figures \ref{fig:bs1} and \ref{fig:bs2}
we have not found imaginary eigenvalues even with 50
basis states, while there they occur already with 25 basis states).
We might point out though  that the pseudo-coulombic basis functions
we used here
are much more suitable for  the
 description of hadronic systems than harmonic oscillator
basis functions used in \cite{bib:parramore}.

We also confirm that there are
problems with  Regge trajectories with scalar confinement in the
Salpeter equation as already found in \cite{bib:lagae2,bib:gara}.
With large quark masses and small number of basis states we show
in Figure \ref{fig:rs}
 that Regge trajectories are not linear.
The situation is not improved even in the heavy-light limit,
as can be seen in Figure \ref{fig:rs2}. We would like to point
out that this can be easily understood from the so called
``no-pair'' equation. The Salpeter equation and its reduced version
in the heavy-light limit are the same
as the no-pair equation \cite{bib:long}, and for the
no-pair equation with scalar confinement
it was shown analytically \cite{bib:ovw}
that linear Regge behavior is lost. If the linear
 Regge behavior is lost in the heavy-light limit, one cannot expect
that it will be restored when both quarks have finite mass.

Finally, let us just mention that we have also investigated
the stability of a mixture of  time component vector and
scalar confinement, i.e. for the kernels of the type
\beq
x \gamma^{0}\otimes \gamma^{0} + (1-x)\bone\otimes\bone\ .
\label{eq:mix}
\eeq
This type of confining kernel (with $x=0.5$), together with a
one gluon exchange kernel,
 was recently used in \cite{bib:munz3} for the investigation
of the weak decays of $B$ and $D$ mesons.
In order to
illustrate this type of confinement we show
in the Figure \ref{fig:mix}
 what happens as $x$ goes from 0.49 to 0.51
(with zero mass quarks and pure confining potential).
Obviously, in this case solutions are stable only if $x>0.5$.
With the addition of a short range  Coulomb
potential, variational solutions exist also for $x=0.5$. A similar
conclusion was also reached in
\cite{bib:parramore}.

\subsection{$\gamma^{\mu}\otimes\gamma_{\mu}$ kernel [full Lorentz
vector]}

Full vector confinement behaves even worse than scalar confinement,
as far as the variational method is concerned. As one can see
in Figure \ref{fig:bv}, the calculation is not stable even
with quark masses as high as $5.0\ GeV$. Increasing the number of
the basis states only makes things worse, as well as decreasing
the quark mass. It is also interesting to note that
this kernel exhibits similar problems even with a
pure Coulomb potential, which was, on the other hand,
found to be stable with scalar and time component vector
kernels. Imaginary eigenvalues with this type
of kernel are quite common, but again we emphasize that
all such solutions have zero norm and must  be rejected.

\section{Conclusions}
\label{sec:con}

In this paper we have corrected a small inconsistency
in Laga$\ddot{\rm e}$'s derivation of the
radial equations for the full Salpeter equation \cite{bib:lagae1},
and extended his analysis to the case of a full Lorentz vector
kernel.
We have concentrated here on the nature of variational solutions
to the full Salpeter equation with a linear confining
potential and three different types of kernels: time component
vector, full vector and Lorentz scalar.
In each case we have examined the stability of
variational solutions, and, when possible, the Regge
structure in the equal mass and the heavy-light cases.
Our results support previous conclusions that scalar
confinement yields unstable variational solutions \cite{bib:parramore}
and non-linear Regge trajectories \cite{bib:lagae2,bib:gara}, even
in the heavy-light limit. On the other hand,
the variational solutions for the time component vector confinement
are stable, and give linear Regge trajectories
with the expected slopes for both the equal mass and heavy-light cases.
In addition we have found that variational
solutions for full vector confinement are even
more unstable than the ones for  scalar confinement.
We emphasize that our analysis took into account
the norm of the solutions.
Our numerical results completely support the theoretical
conclusions of \cite{bib:lagae2,bib:munz2} that
eigenvalues for the states with physical norms are
always real. Imaginary eigenvalues do appear,
and are quite common for  full vector confinement,
but these solutions
always have zero norm.

The occurrence of non-linear Regge trajectories in scalar confinement
is an extension of the results of Gara et. al. \cite{bib:gara}
(for the reduced  Salpeter equation) and  Laga$\ddot{\rm e}$
\cite{bib:lagae2} (for the full Salpeter equation) in the
equal mass case. We have found that the same problem
persists even in the heavy-light case. The origin of this effect
can be explained analytically from the
so called ``no-pair'' equation in  coordinate space \cite{bib:ovw}.

Even though the time component vector interaction behaves perfectly
in the full Salpeter equation, and leads to the
expected linear Regge trajectories, it cannot be
directly applied in any realistic meson model, because it
conflicts with QCD. The most evident
example of this conflict is its wrong sign of the
spin orbit splitting. It is important
to point out that the relativistic flux tube
model reduces to a time component vector interaction
for $S$-waves, but nevertheless yields the correct
spin-orbit splitting relativistic correction \cite{bib:ovw2}.

\appendix

\newpage

\begin{center}
APPENDICES
\end{center}

\section{Radial equations}
\label{app:eqs}

In this appendix we give the complete set of radial equations
for the kernels considered in this paper. These equations represent
the general case of a quark with mass $m_{1}$ and an
anti-quark with mass $m_{2}$. One has to keep in mind
that for $J=0$ four wave functions vanish, i.e.
 we have $N_{10}=N_{20}=0$,
and $n_{1+}=n_{2+}=0$.

In the equal mass case these equations somewhat simplify, since
one has $E_{1}=E_{2}$, $\phi=\phi_{1}=\phi_{2}$,
and $\theta = 0$, so that
$\stheta=0$ and $\ctheta=1$. Also, since charge conjugation
 is a good quantum number
in the equal mass case, the four $P=(-1)^{J+1}$ state
equations split into two systems of two equations, one corresponding
to $C=(-1)^{J}$ (involving $L_{1}$ and $L_{2}$), and the other
corresponding to $C=(-1)^{J+1}$ (involving $N_{10}$ and $N_{20}$).

The heavy-light limit
($m_{2}\rar \infty$) is obtained by setting $E_{2}\rar m_{2}$,
$\phi_{2}\rar \frac{\pi}{2}$, so that $\stheta\rar \cphi$ and
$\ctheta\rar \sphi$. It is interesting to note that in the heavy-light
limit (and for kernels $\gamma^{0}\otimes\gamma^{0}$ and
$\bone\otimes \bone$, but not for $\gamma^{\mu}\otimes\gamma_{\mu}$ kernel)
physical solutions satisfy $L_{1}=L_{2}$ and $ N_{10}=N_{20}$ (for
the $P=(-1)^{J+1}$ states), or $n_{1+}=n_{2+}$ and $ n_{1-}=n_{2-}$
 (for the $P=(-1)^{J}$ states). Therefore, in these cases the
system of four equations can be  reduced to a system of
only two radial equations.

 Of course, for any mixture
of different kernels, only the kernel parts of
the radial equations should
be added, and the kinetic energy terms are always the same. In the
$\bone\otimes \bone$ case, we have introduced an additional minus sign
in the kernel, so that
$V(r)$ has the same form for all three cases considered, e.g. for the
Cornell potential $V(r) = a r  -\frac{\kappa}{r}$.

\subsection{$\gamma^{0}\otimes\gamma^{0}$ kernel}

States with parity $P=(-1)^{J+1}$:
\bea
M L_{1} &=& [E_{1}+E_{2}] L_{2}
+ \int_{0}^{\infty}
 \frac{k'^{2}dk'}{(2\pi)^{2}} [\ctheta V_{J} \ctheta' L_{2}'
\nonumber \\
&+&\stheta (\mu^{2}V_{J-1}+\nu^{2}V_{J+1})\stheta'L_{2}'
+\mu\nu \stheta (V_{J-1}-V_{J+1})\cphi' N_{20}'] \ ,\nonumber\\
M L_{2} &=& [E_{1}+E_{2}] L_{1}
+ \int_{0}^{\infty}
 \frac{k'^{2}dk'}{(2\pi)^{2}} [\sphi V_{J} \sphi' L_{1}'
\nonumber \\
&+&\cphi (\mu^{2}V_{J-1}+\nu^{2}V_{J+1})\cphi'L_{1}'
+\mu\nu \cphi (V_{J-1}-V_{J+1})\stheta' N_{10}'] \ ,\\
M N_{10} &=& [E_{1}+E_{2}] N_{20}
+ \int_{0}^{\infty}
 \frac{k'^{2}dk'}{(2\pi)^{2}} [\sphi V_{J} \sphi' N_{20}'
\nonumber \\
&+&\cphi (\nu^{2}V_{J-1}+\mu^{2}V_{J+1})\cphi'N_{20}'
+\mu\nu \cphi (V_{J-1}-V_{J+1})\stheta' L_{2}'] \ ,\nonumber\\
M N_{20} &=& [E_{1}+E_{2}] N_{10}
+ \int_{0}^{\infty}
 \frac{k'^{2}dk'}{(2\pi)^{2}} [\ctheta V_{J} \ctheta' N_{10}'
\nonumber \\
&+&\stheta (\nu^{2}V_{J-1}+\mu^{2}V_{J+1})\stheta'N_{10}'
+\mu\nu \stheta (V_{J-1}-V_{J+1})\cphi' L_{1}'] \ .\nonumber
\eea

States with parity $P=(-1)^{J}$:
\bea
M n_{1+} &=& [E_{1}+E_{2}] n_{2+}
+ \int_{0}^{\infty}
 \frac{k'^{2}dk'}{(2\pi)^{2}} [\cphi V_{J} \cphi' n_{2+}'
\nonumber \\
&+&\sphi (\nu^{2}V_{J-1}+\mu^{2}V_{J+1})\sphi'n_{2+}'
+\mu\nu \sphi (V_{J-1}-V_{J+1})\ctheta' n_{2-}'] \ ,\nonumber\\
M n_{2+} &=& [E_{1}+E_{2}] n_{1+}
+ \int_{0}^{\infty}
 \frac{k'^{2}dk'}{(2\pi)^{2}} [\stheta V_{J} \stheta' n_{1+}'
\nonumber \\
&+&\ctheta (\nu^{2}V_{J-1}+\mu^{2}V_{J+1})\ctheta'n_{1+}'
+\mu\nu \ctheta (V_{J-1}-V_{J+1})\sphi' n_{1-}'] \ ,\\
M n_{1-} &=& [E_{1}+E_{2}] n_{2-}
+ \int_{0}^{\infty}
 \frac{k'^{2}dk'}{(2\pi)^{2}} [\stheta V_{J} \stheta' n_{2-}'
\nonumber \\
&+&\ctheta (\mu^{2}V_{J-1}+\nu^{2}V_{J+1})\ctheta'n_{2-}'
+\mu\nu \ctheta (V_{J-1}-V_{J+1})\sphi' n_{2+}'] \ ,\nonumber\\
M n_{2-} &=& [E_{1}+E_{2}] n_{1-}
+ \int_{0}^{\infty}
 \frac{k'^{2}dk'}{(2\pi)^{2}} [\cphi V_{J} \cphi' n_{1-}'
\nonumber \\
&+&\sphi (\mu^{2}V_{J-1}+\nu^{2}V_{J+1})\sphi'n_{1-}'
+\mu\nu \sphi (V_{J-1}-V_{J+1})\ctheta' n_{1+}'] \ .\nonumber
\eea

\subsection{$\bone\otimes \bone$ kernel}

States with parity $P=(-1)^{J+1}$:
\bea
M L_{1} &=& [E_{1}+E_{2}] L_{2}
+ \int_{0}^{\infty}
 \frac{k'^{2}dk'}{(2\pi)^{2}} [\ctheta V_{J} \ctheta' L_{2}'
\nonumber \\
&-&\stheta (\mu^{2}V_{J-1}+\nu^{2}V_{J+1})\stheta'L_{2}'
-\mu\nu \stheta (V_{J-1}-V_{J+1})\cphi' N_{20}'] \ ,\nonumber\\
M L_{2} &=& [E_{1}+E_{2}] L_{1}
+ \int_{0}^{\infty}
 \frac{k'^{2}dk'}{(2\pi)^{2}} [\sphi V_{J} \sphi' L_{1}'
\nonumber \\
&-&\cphi (\mu^{2}V_{J-1}+\nu^{2}V_{J+1})\cphi'L_{1}'
-\mu\nu \cphi (V_{J-1}-V_{J+1})\stheta' N_{10}'] \ ,\\
M N_{10} &=& [E_{1}+E_{2}] N_{20}
+ \int_{0}^{\infty}
 \frac{k'^{2}dk'}{(2\pi)^{2}} [\sphi V_{J} \sphi' N_{20}'
\nonumber \\
&-&\cphi (\nu^{2}V_{J-1}+\mu^{2}V_{J+1})\cphi'N_{20}'
-\mu\nu \cphi (V_{J-1}-V_{J+1})\stheta' L_{2}'] \ ,\nonumber\\
M N_{20} &=& [E_{1}+E_{2}] N_{10}
+ \int_{0}^{\infty}
 \frac{k'^{2}dk'}{(2\pi)^{2}} [\ctheta V_{J} \ctheta' N_{10}'
\nonumber \\
&-&\stheta (\nu^{2}V_{J-1}+\mu^{2}V_{J+1})\stheta'N_{10}'
-\mu\nu \stheta (V_{J-1}-V_{J+1})\cphi' L_{1}'] \ .\nonumber
\eea

States with parity $P=(-1)^{J}$:
\bea
M n_{1+} &=& [E_{1}+E_{2}] n_{2+}
+ \int_{0}^{\infty}
 \frac{k'^{2}dk'}{(2\pi)^{2}} [-\cphi V_{J} \cphi' n_{2+}'
\nonumber \\
&+&\sphi (\nu^{2}V_{J-1}+\mu^{2}V_{J+1})\sphi'n_{2+}'
+\mu\nu \sphi (V_{J-1}-V_{J+1})\ctheta' n_{2-}'] \ ,\nonumber\\
M n_{2+} &=& [E_{1}+E_{2}] n_{1+}
+ \int_{0}^{\infty}
 \frac{k'^{2}dk'}{(2\pi)^{2}} [-\stheta V_{J} \stheta' n_{1+}'
\nonumber \\
&+&\ctheta (\nu^{2}V_{J-1}+\mu^{2}V_{J+1})\ctheta'n_{1+}'
+\mu\nu \ctheta (V_{J-1}-V_{J+1})\sphi' n_{1-}'] \ ,\\
M n_{1-} &=& [E_{1}+E_{2}] n_{2-}
+ \int_{0}^{\infty}
 \frac{k'^{2}dk'}{(2\pi)^{2}} [-\stheta V_{J} \stheta' n_{2-}'
\nonumber \\
&+&\ctheta (\mu^{2}V_{J-1}+\nu^{2}V_{J+1})\ctheta'n_{2-}'
+\mu\nu \ctheta (V_{J-1}-V_{J+1})\sphi' n_{2+}'] \ ,\nonumber\\
M n_{2-} &=& [E_{1}+E_{2}] n_{1-}
+ \int_{0}^{\infty}
 \frac{k'^{2}dk'}{(2\pi)^{2}} [-\cphi V_{J} \cphi' n_{1-}'
\nonumber \\
&+&\sphi (\mu^{2}V_{J-1}+\nu^{2}V_{J+1})\sphi'n_{1-}'
+\mu\nu \sphi (V_{J-1}-V_{J+1})\ctheta' n_{1+}'] \ .\nonumber
\eea

\subsection{$\gamma^{\mu}\otimes\gamma_{\mu}$ kernel}

States with parity $P=(-1)^{J+1}$:
\bea
M L_{1} &=& [E_{1}+E_{2}] L_{2}
+ \int_{0}^{\infty}
 \frac{k'^{2}dk'}{(2\pi)^{2}} [4\ctheta V_{J} \ctheta' L_{2}'
\nonumber \\
&+&2\stheta (\mu^{2}V_{J-1}+\nu^{2}V_{J+1})\stheta'L_{2}'
+2\mu\nu \stheta (V_{J-1}-V_{J+1})\cphi' N_{20}'] \ ,\nonumber\\
M L_{2} &=& [E_{1}+E_{2}] L_{1}
+ \int_{0}^{\infty}
 \frac{k'^{2}dk'}{(2\pi)^{2}} [-2\sphi V_{J} \sphi' L_{1}']\ ,
 \\
M N_{10} &=& [E_{1}+E_{2}] N_{20}
+ \int_{0}^{\infty}
 \frac{k'^{2}dk'}{(2\pi)^{2}} [
2\cphi (\nu^{2}V_{J-1}+\mu^{2}V_{J+1})\cphi'N_{20}'
\nonumber \\
&+&2\mu\nu \cphi (V_{J-1}-V_{J+1})\stheta' L_{2}'] \ ,\nonumber\\
M N_{20} &=& [E_{1}+E_{2}] N_{10}
+ \int_{0}^{\infty}
 \frac{k'^{2}dk'}{(2\pi)^{2}} [2\ctheta V_{J} \ctheta' N_{10}']\ .
\nonumber
\eea

States with parity $P=(-1)^{J}$:
\bea
M n_{1+} &=& [E_{1}+E_{2}] n_{2+}
+ \int_{0}^{\infty}
 \frac{k'^{2}dk'}{(2\pi)^{2}} [2\cphi V_{J} \cphi' n_{2+}']\ ,
\nonumber \\
M n_{2+} &=& [E_{1}+E_{2}] n_{1+}
+ \int_{0}^{\infty}
 \frac{k'^{2}dk'}{(2\pi)^{2}} [
2\ctheta (\nu^{2}V_{J-1}+\mu^{2}V_{J+1})\ctheta'n_{1+}'\nonumber \\
&+&2\mu\nu \ctheta (V_{J-1}-V_{J+1})\sphi' n_{1-}'] \ ,\\
M n_{1-} &=& [E_{1}+E_{2}] n_{2-}
+ \int_{0}^{\infty}
 \frac{k'^{2}dk'}{(2\pi)^{2}} [-2\stheta V_{J} \stheta' n_{2-}']\ ,
\nonumber \\
M n_{2-} &=& [E_{1}+E_{2}] n_{1-}
+ \int_{0}^{\infty}
 \frac{k'^{2}dk'}{(2\pi)^{2}} [4\cphi V_{J} \cphi' n_{1-}'
\nonumber \\
&+&2\sphi (\mu^{2}V_{J-1}+\nu^{2}V_{J+1})\sphi'n_{1-}'
+2\mu\nu \sphi (V_{J-1}-V_{J+1})\ctheta' n_{1+}'] \ .\nonumber
\eea

\section{Numerical solution of the radial equations}
\label{app:ns}

As briefly mentioned in Section \ref{sec:res}, the
 easiest way to solve the
radial equations given in Appendix \ref{app:eqs}
is to expand the radial wave functions $f(k)$ in terms
of some complete set of basis functions $\{e_{iL}(k)\}$, and truncate
the expansion to the first $N$ basis states, i.e.
\beq
f(k)\simeq \sum_{i=0}^{N-1} c_{i}^{f}e_{iL}(k)\ .
\eeq
In this way a set of $n$  coupled radial equations becomes the
$nN\times nN$ matrix equation
\beq
\cH \psi = M\psi\ ,
\eeq
where $\psi$ is an $nN$ dimensional vector,
\beq
\psi = \left(
\begin{array}{c}
f_{1} \\
\vdots \\
f_{n} \end{array}
\right)\ .
\eeq
The matrix $\cH$ and its eigenvalues depend
 on the variational parameter $\beta$
characterizing the basis functions. However, if the calculation
is stable, dependence of the solution on $\beta$ should
reduce as $N$  increases. This is manifested
by the development of plateaus in $\beta$ having the same
eigenvalues.

A basis set that was shown to be very successful in calculations
of this sort is given by \cite{bib:weniger}
\beq
e_{iL}(\beta,k) = (-i)^{L}\tilde{N}_{iL}\beta^{\frac{1}{2}}
(\frac{\beta}{\beta^{2}+k^{2}})^{L+2}k^{L}
P_{i}^{(L+\frac{3}{2},L+\frac{1}{2})}
(\frac{k^{2}-\beta^{2}}{k^{2}+\beta^{2}})\ ,\label{eq:ep}
\eeq
where $P_{i}^{(a,b)}(x)$ are Jacobi polynomials and
\beq
\tilde{N}_{iL}=\frac{2\Gamma(\frac{1}{2})}{\Gamma (i+L+\frac{3}{2})}
\left[\frac{i!(i+2L+2)!}{\pi}\right]^{\frac{1}{2}}\ .
\eeq
The Fourier transform of (\ref{eq:ep}) is known analytically,
\beq
e_{iL}(\beta,r) = N_{iL}\beta^{\frac{3}{2}}(2\beta r)^{L}
e^{-\beta r} L_{i}^{(2L+2)}(2\beta r)\ ,\label{eq:er}
\eeq
with $L_{i}^{(a)}(x)$ being the generalized Laguerre polynomials and
\beq
N_{iL} = \left[\frac{8(i!)}{(i+2L+2)!}\right]^{\frac{1}{2}}\ .
\eeq

The kinetic energy terms of the $\cH$  matrix
(dropping the dependence of the basis states on $\beta$),
\beq
\int_{0}^{\infty} k^{2}dk e_{iL}^{*}(k)\sqrt{k^{2}+m^{2}}e_{jL}(k)\ ,
\eeq
can be efficiently calculated using Gauss-Jacobi
quadrature formula after performing a change of integration
variable to $x=\frac{k^{2}-\beta^{2}}{k^{2}+\beta^{2}}$.
Keeping in mind the definition (\ref{eq:vdef}),
the kernel part of $\cH$ will in general include terms like
\beq
\frac{2}{\pi}\int_{0}^{\infty} k^{2}dk \int_{0}^{\infty} k'^{2}dk'
\int_{0}^{\infty} r^{2}dr e_{iL}^{*}(k)F(k)j_{L'}(kr)V(r)
j_{L'}(k'r)G(k')e_{jL}(k')\ ,\label{eq:kp}
\eeq
where $F$ and $G$ are some functions of $k$ and $k'$, respectively.
It is quite difficult to approximate these integrals by applying
standard numerical quadrature methods, since the
range of integration is infinite and spherical Bessel functions
$j_{l}(kr)$ are rapidly oscillating. Besides, we would have to perform
the quadrature for each value of $r$ where $V(r)j_{L'}(kr)j_{L'}(k'r)$
is needed. Therefore, it is much better to expand
$F(k)e_{iL}(k)$ in terms of basis functions
$\{e_{i'L'}\}$ \cite{bib:jacobsth}, i.e.
\beq
F(k)e_{iL}(k) \simeq \sum_{i'=0}^{N'-1}
c^{(F,L',L)}_{i'i}e_{i'L'}(k)\ ,
\eeq
where $N'$ is the number of basis states of $\{e_{i'L'}\}$
used (similarly we have to expand $G(k')e_{jL}(k')$). Now, the
radial  Fourier
transforms,
\beq
e_{iL}(r) = i^{L}\frac{4\pi}{(2\pi)^{\frac{3}{2}}}
\int_{0}^{\infty} k^{2}dk j_{L}(kr) e_{iL}(k)\ ,
\eeq
allow us to perform integrations over $k$ and $k'$
analytically, so that (\ref{eq:kp}) becomes
\beq
\sum_{i'=0}^{N'-1}\sum_{j'=0}^{N'-1}c^{(F,L',L)*}_{i'i}
V_{i'j'}^{(L')}c^{(G,L',L)}_{j'j}\ ,
\eeq
where,
\bea
c^{(F,L',L)}_{i'i} &=& \int_{0}^{\infty} k^{2}dk
e_{i'L'}^{*}(k)F(k)e_{iL}(k)\ ,\\
V_{i'j'}^{(L')}&=&\int_{0}^{\infty} r^{2}dr e_{i'L'}^{*}(r)V(r) e_{j'L'}(r)\ .
\eea

Again, the integrals involved in the
calculation of $c^{(F,L,L')}_{i'i}$
can be efficiently evaluated using the Gauss-Jacobi quadrature formula,
while matrix elements of $V(r)$ were calculated analytically
\cite{bib:ovw} for the short range Coulomb and linear confining
potential that were used in this paper.

In practice, we choose $L$ in such a way that $L=J$ for the
states with parity $P=(-1)^{J+1}$, and $L=J-1$ for the
states with parity $P=(-1)^{J}$ (unless $J=0$, when we
take $L=1$. From the radial equations given in Appendix \ref{app:eqs}
one can see that $L'$ can take values
$J-1$, $J$, or $J+1$, and we have found that generally
taking $N'=2N$ is more than enough
to accurately describe all functions of the form $F(k)e_{iL}(k)$.

In order to make  sure that numerical calculations
were done correctly, we have written two independent
programs, each of them treating the general (unequal mass) case,
and also two limiting cases (equal mass and heavy-light limit) separately,
and for all kernels considered in the text. Besides consistency of
the two programs of ours, and consistency
of the two special cases with the general case, we have also checked our
results against those of \cite{bib:lagae2}, where only equal mass
case was considered, and of \cite{bib:parramore}, where stability of
pseudoscalar case was considered.

\vskip 1cm
\begin{center}
ACKNOWLEDGMENTS
\end{center}
We thank Jean-Francois Laga$\ddot{\rm e}$ and Loyal Durand
for helpful comments.
This work was supported in part by the U.S. Department of Energy
under Contract Nos.  DE-FG02-95ER40896 and DE-AC05-84ER40150,
the National Science
Foundation under Grant No. HRD9154080,
and in part by the University
of Wisconsin Research Committee with funds granted by the Wisconsin Alumni
Research Foundation.

\newpage

\begin{figure}
\begin{center}
FIGURES
\end{center}
\vskip 0.2cm
\caption{The lowest three states with positive norm for
time component vector confinement ($V(r)=ar$, $a = 0.2\ GeV^{2}$) in the
pseudoscalar case, with
$m_{1}=m_{2}=0$. Calculations were done using 5, 15, 25 and 50
basis states.}
\label{fig:bv0}
\end{figure}

\begin{figure}
\caption{Regge trajectories for a time component vector confinement
($V(r)=ar$, $a=0.2\ GeV^{2}$)
in the light-light case with
 equal mass quarks $P=(-1)^{J+1}$ and
$C=(-1)^{J}$. We have taken $m_{1}=m_{2}=0$,
and used 25 basis states.}
\label{fig:rv0}
\end{figure}

\begin{figure}
\caption{Regge trajectories for a time component
vector confinement ($V(r)=ar$, $a=0.2\ GeV^{2}$)
in the heavy-light case with $P=(-1)^{J+1}$.
We have taken $m_{1}=0$,
$m_{2}=5.0\ GeV$,
and used 25 basis states in the
equations corresponding to the heavy-light limit.}
\label{fig:r2v0}
\end{figure}

\begin{figure}
\caption{Radial wave functions in  coordinate
space for the $J^{PC}=0^{-+}$ ($^{1}S_{0}$ state, $L_{1}$
is the lower, and $L_{2}$ the upper full line), and
$J^{PC}=0^{++}$ ($^{3}P_{0}$ state, $n_{1-}$
is the upper, and $n_{2-}$ the lower dashed line),
with
a time component vector kernel and $V(r)=ar$
($a=0.2\ GeV^{2}$).
We have chosen
quark masses $m_{1}=m_{2}=0$. The calculation was
done with 15 basis states.}
\label{fig:ff}
\end{figure}

\begin{figure}
\caption{Comparison
of the Coulombic
ground  state energy
as a function of light quark mass $m_{1}$, obtained
from Dirac the (full lines) and heavy-light Salpeter equation (dashed
lines),
with a time component vector Coulomb potential
($V(r)=-\frac{\kappa}{r}$).}
\label{fig:dv0}
\end{figure}

\begin{figure}
\caption{The lowest three states with positive norm for
scalar confinement ($V(r)=ar$, $a = 0.29\ GeV^{2}$) in the
pseudoscalar case, with
$m_{1}=m_{2}=0.9\ GeV$. Calculations were done using 5, 25 and 35
basis states.}
\label{fig:bs1}
\end{figure}

\begin{figure}
\caption{The lowest pseudoscalar state with positive norm for
scalar confinement ($V(r)=ar$, $a = 0.29\ GeV^{2}$) with
$m_{1}=m_{2}=0.9\ GeV$. Calculations with 39 (dashed line) and
40 (full line) basis states are shown. This illustrates
the onset of instability for this example of scalar confinement.}
\label{fig:bs2}
\end{figure}

\begin{figure}
\caption{Pseudoscalar ground state
radial wave functions $L_{1}$ and $L_{2}$
in  coordinate
space
with
a scalar confining kernel and $V(r)=ar$
($a=0.29\ GeV^{2}$).
Quark masses were $m_{1}=m_{2}=0.9\ GeV$.
Calculations with 39 (dashed line) and
40 (full line) basis states are shown, and
variational parameter was $\beta=1.538\ GeV$.}
\label{fig:ff1}
\end{figure}

\begin{figure}
\caption{Regge trajectories for  scalar confinement ($V(r)=ar$)
with equal mass quarks and $P=(-1)^{J+1}$ and
$C=(-1)^{J}$. We have taken $m_{1}=m_{2}=1\ GeV$,
$a=0.2\ GeV^{2}$, and used 25 basis states.}
\label{fig:rs}
\end{figure}

\begin{figure}
\caption{Regge trajectories for  scalar
confinement ($V(r)=ar$, $a=0.2\ GeV^{2}$)
in the heavy-light case with $P=(-1)^{J+1}$.
We have taken $m_{1}=0.1\ GeV$,
$m_{2}=5.0\ GeV$,
and used 25 basis states in the
equations corresponding to the heavy-light limit.}
\label{fig:rs2}
\end{figure}

\begin{figure}
\caption{The lowest state with positive norm
for the mixture of time component vector and scalar
confinement as given in (\protect\ref{eq:mix})
($V(r)=ar,\ a=0.2GeV^{2},\ m_{1}=m_{2}=0$).
Calculations with $x=0.51$ (full line), $x=0.50$ (dashed line),
and $x=0.49$ (dotted line), correspond to
51\%, 50\%, and 49\% of the time component vector kernel.
We have used 15 basis states.}
\label{fig:mix}
\end{figure}

\begin{figure}
\caption{The lowest state with positive norm for full Lorentz
vector confinement ($V(r)=ar$, $a = 0.2\ GeV^{2}$) in the
pseudoscalar equal mass case, with the choice of
$m_{1}=m_{2}=5.0\ GeV$. Calculations with 5 (dashed line) and
15 (full line) basis states are shown.}
\label{fig:bv}
\end{figure}

\clearpage

\begin{figure}[p]
\epsfxsize = 5.4in \epsfbox{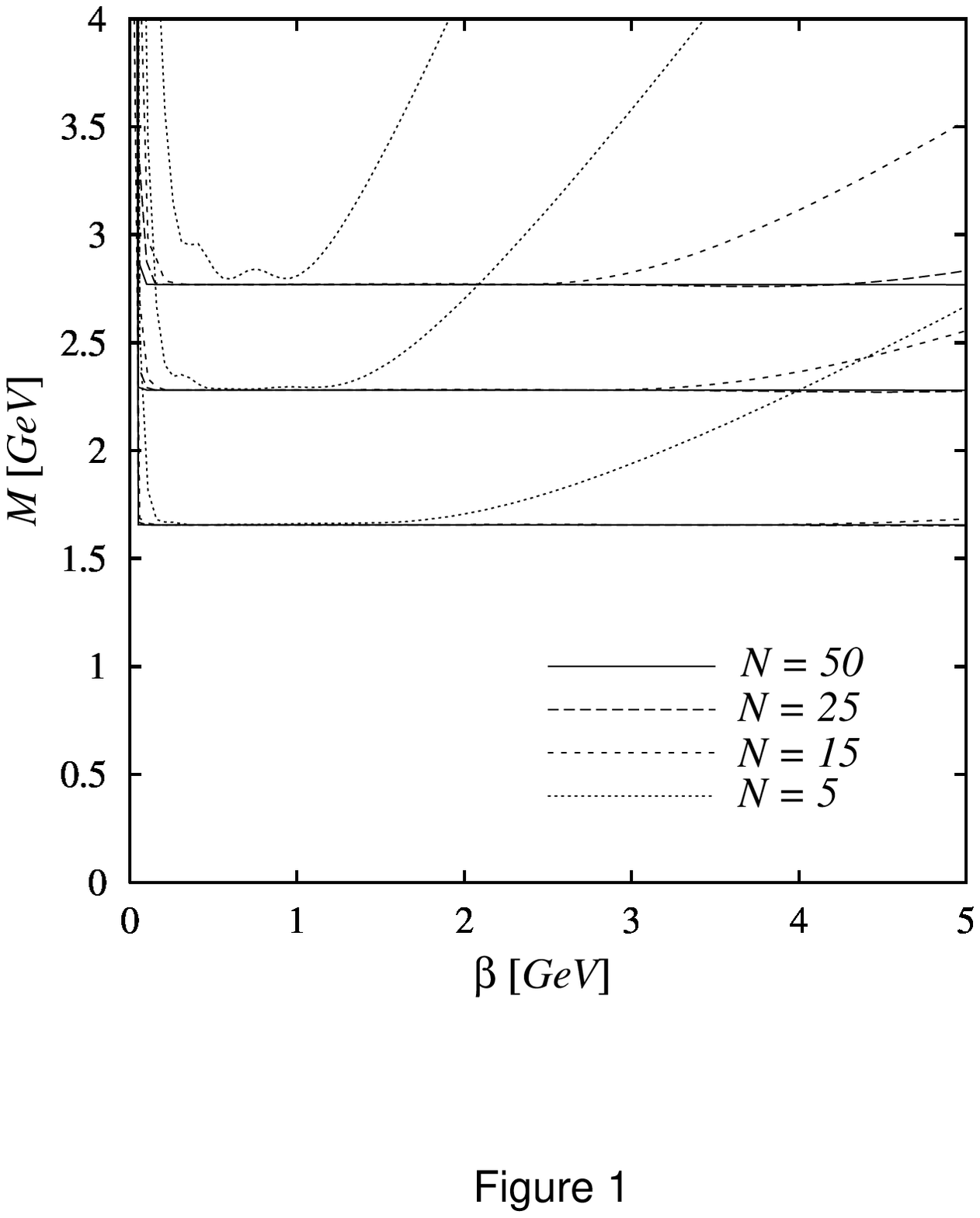}
\end{figure}

\begin{figure}[p]
\epsfxsize = 5.4in \epsfbox{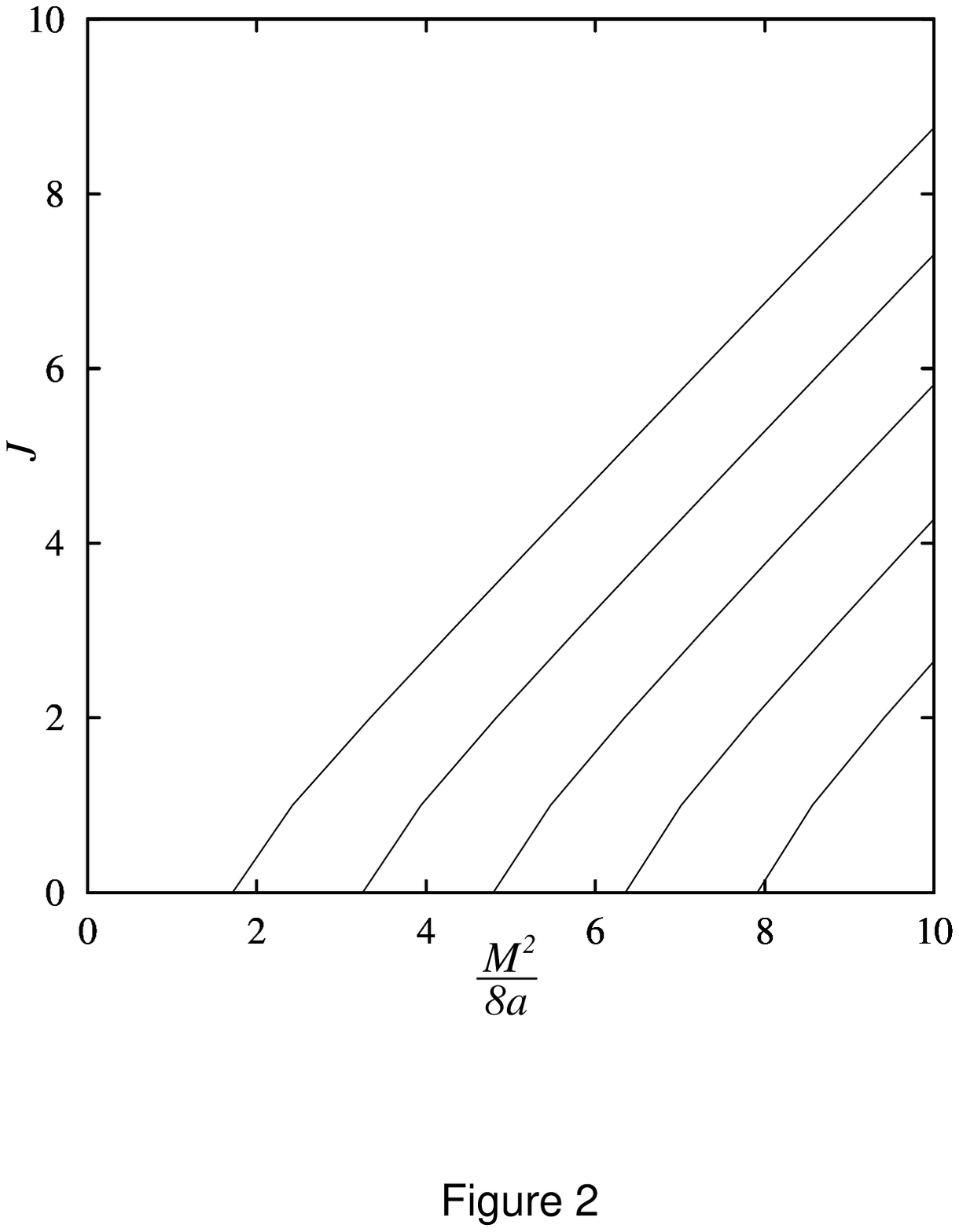}
\end{figure}

\begin{figure}[p]
\epsfxsize = 5.4in \epsfbox{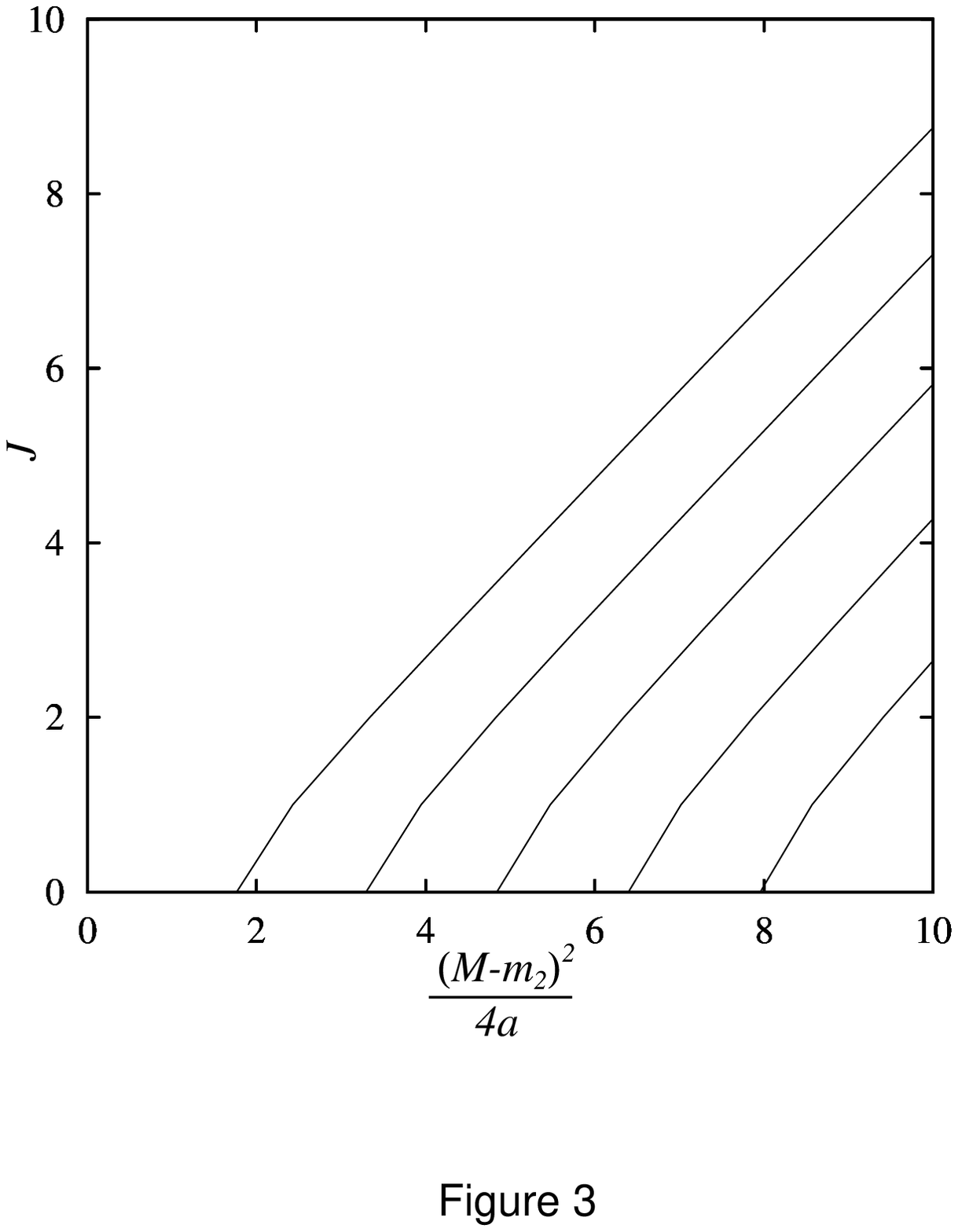}
\end{figure}

\begin{figure}[p]
\epsfxsize = 5.4in \epsfbox{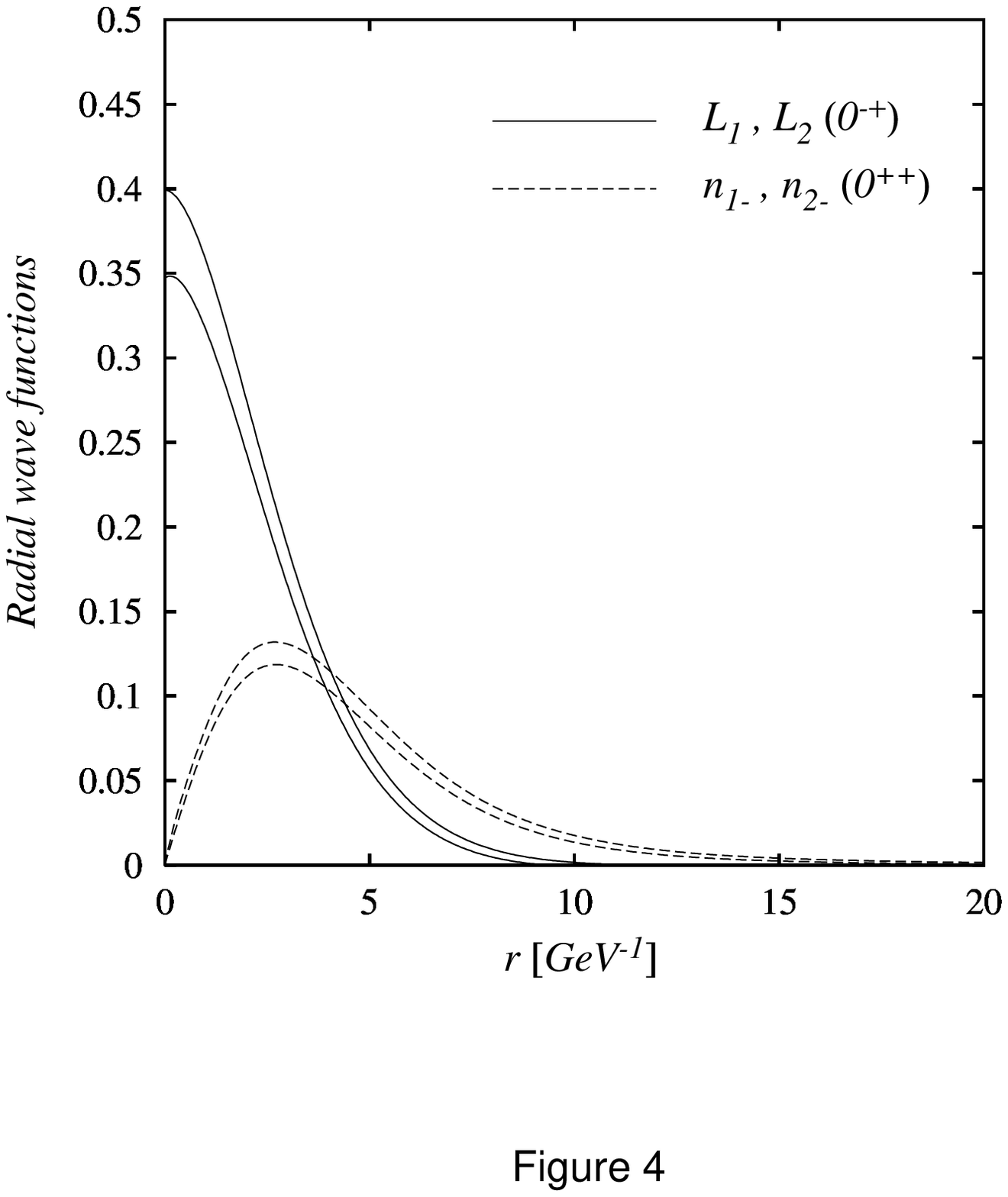}
\end{figure}

\begin{figure}[p]
\epsfxsize = 5.4in \epsfbox{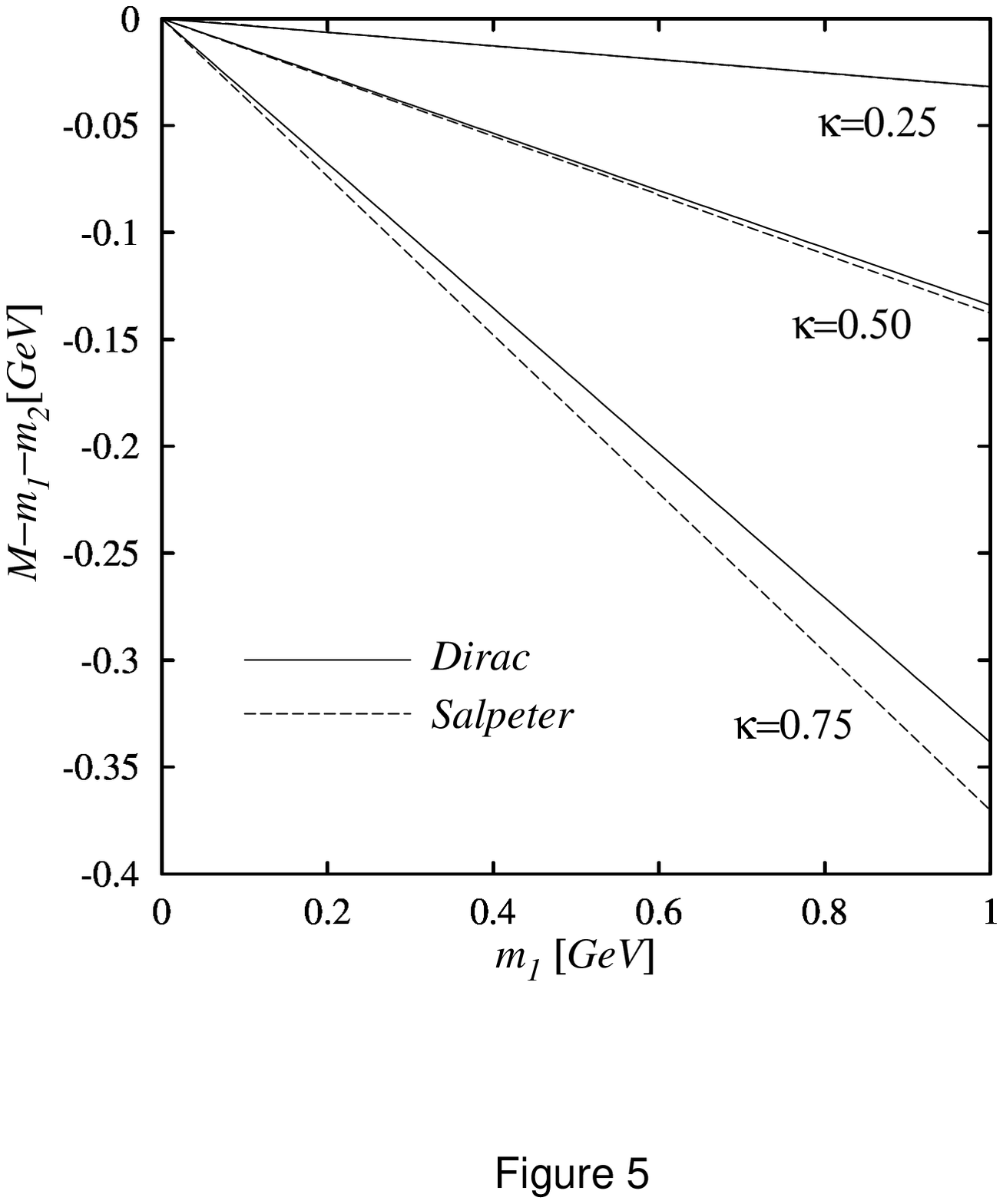}
\end{figure}

\begin{figure}[p]
\epsfxsize = 5.4in \epsfbox{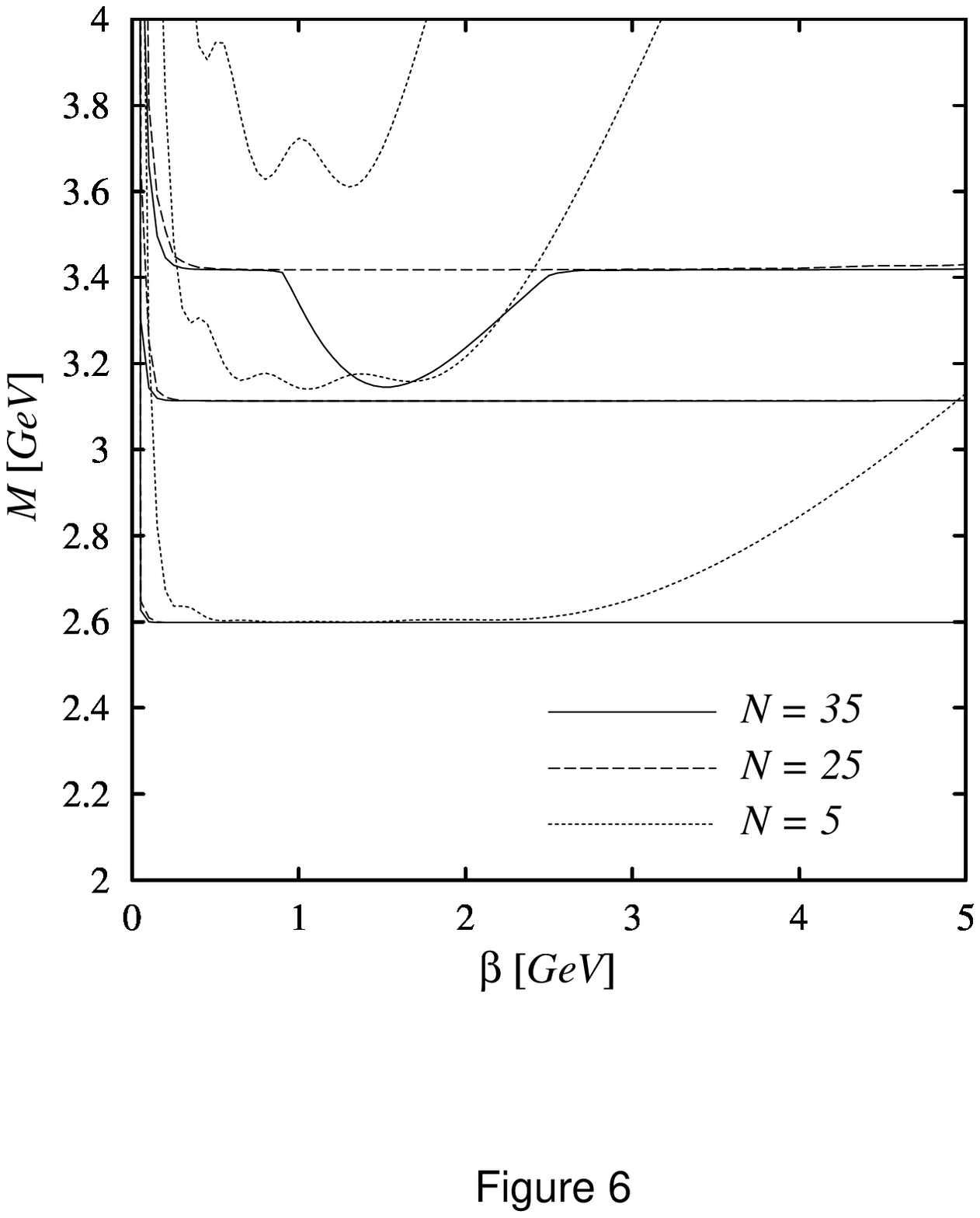}
\end{figure}

\begin{figure}[p]
\epsfxsize = 5.4in \epsfbox{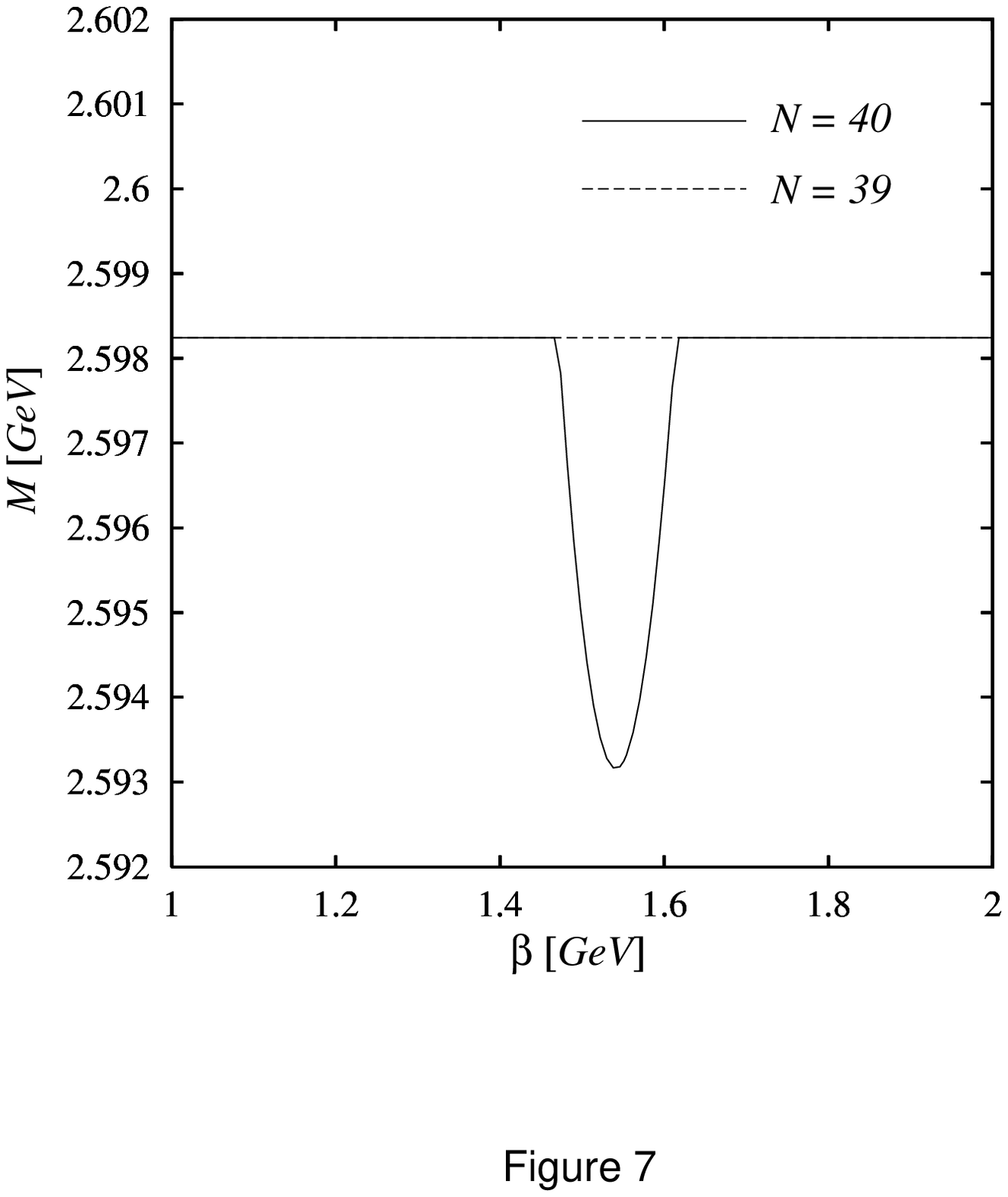}
\end{figure}

\begin{figure}[p]
\epsfxsize = 5.4in \epsfbox{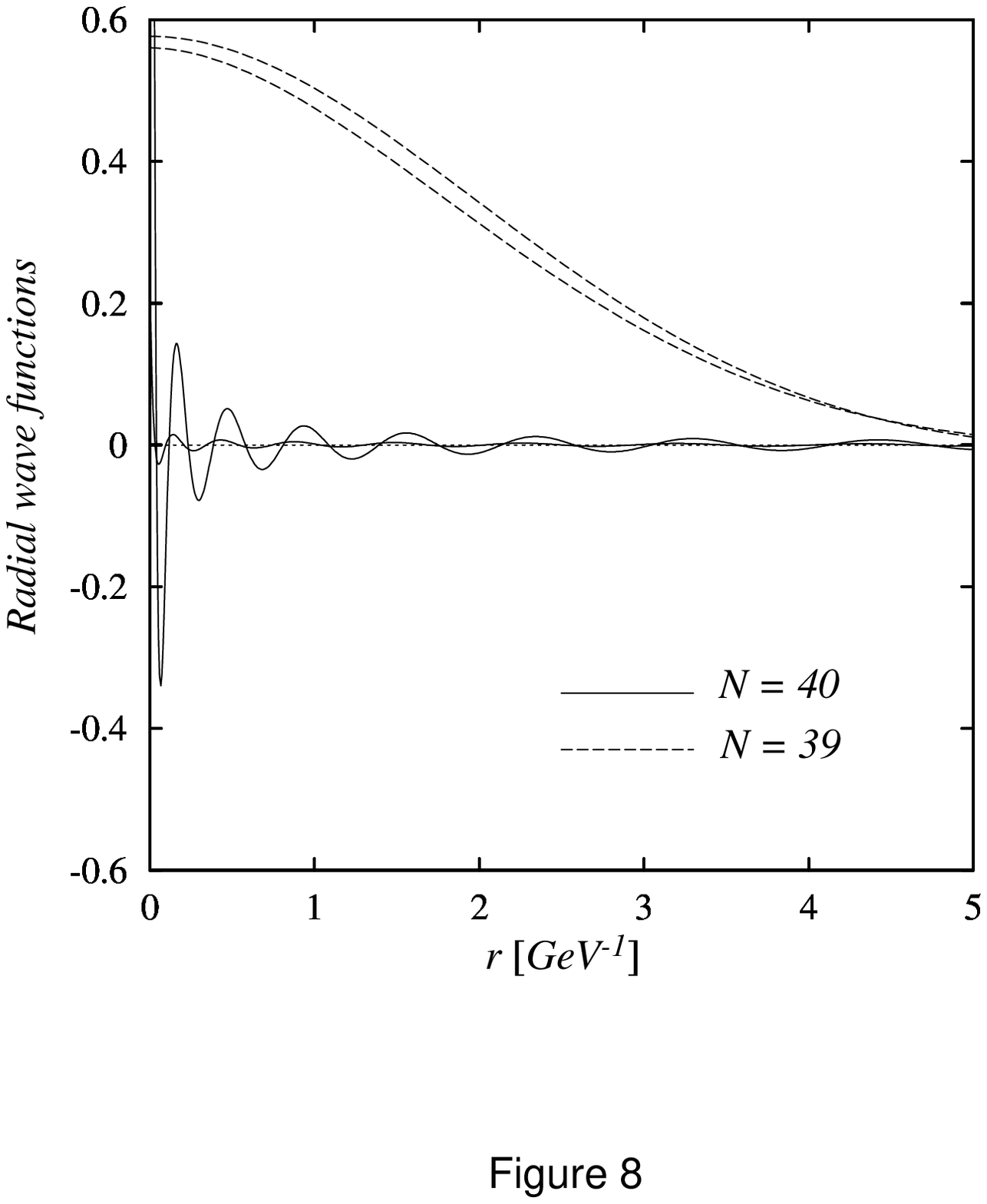}
\end{figure}

\begin{figure}[p]
\epsfxsize = 5.4in \epsfbox{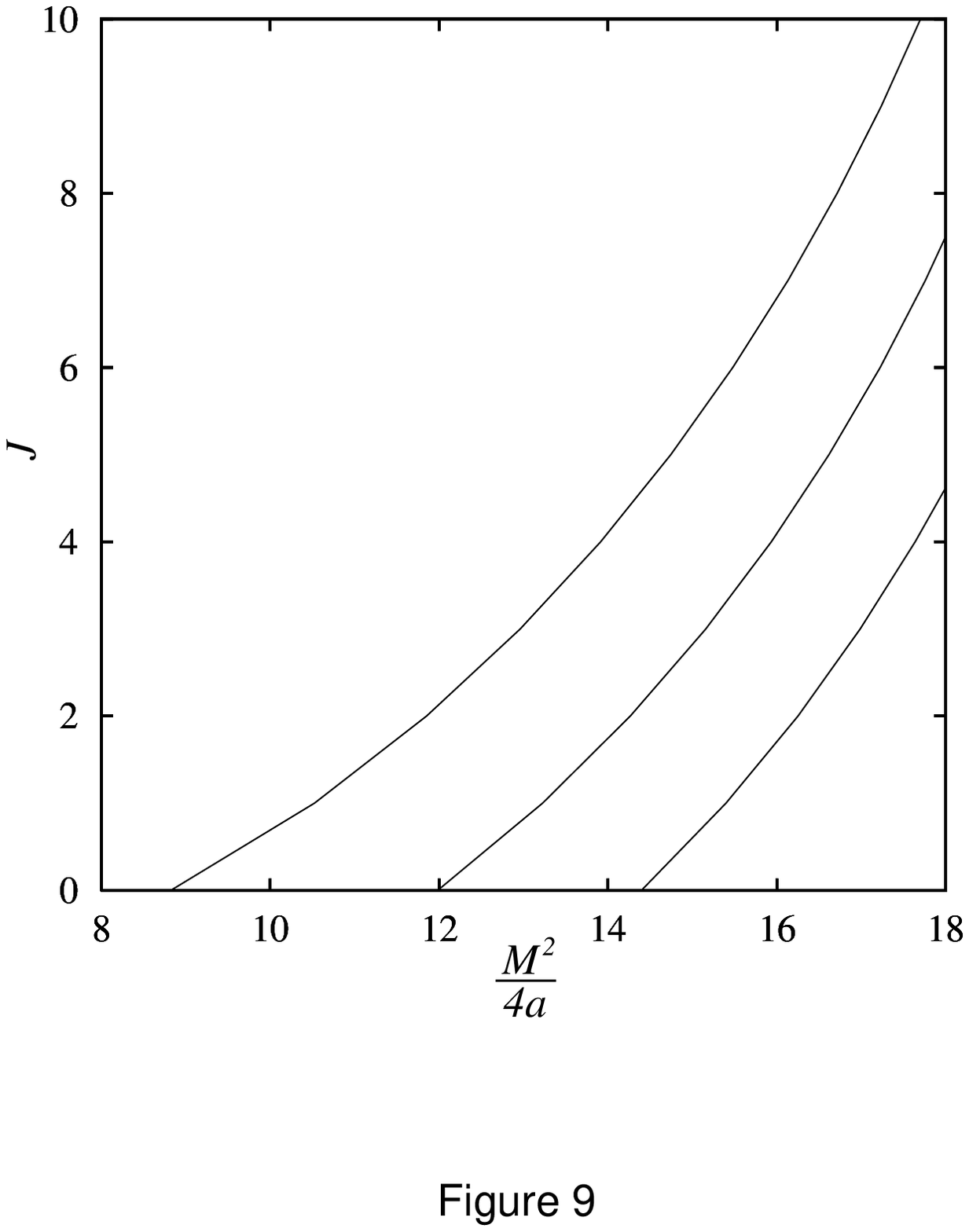}
\end{figure}

\begin{figure}[p]
\epsfxsize = 5.4in \epsfbox{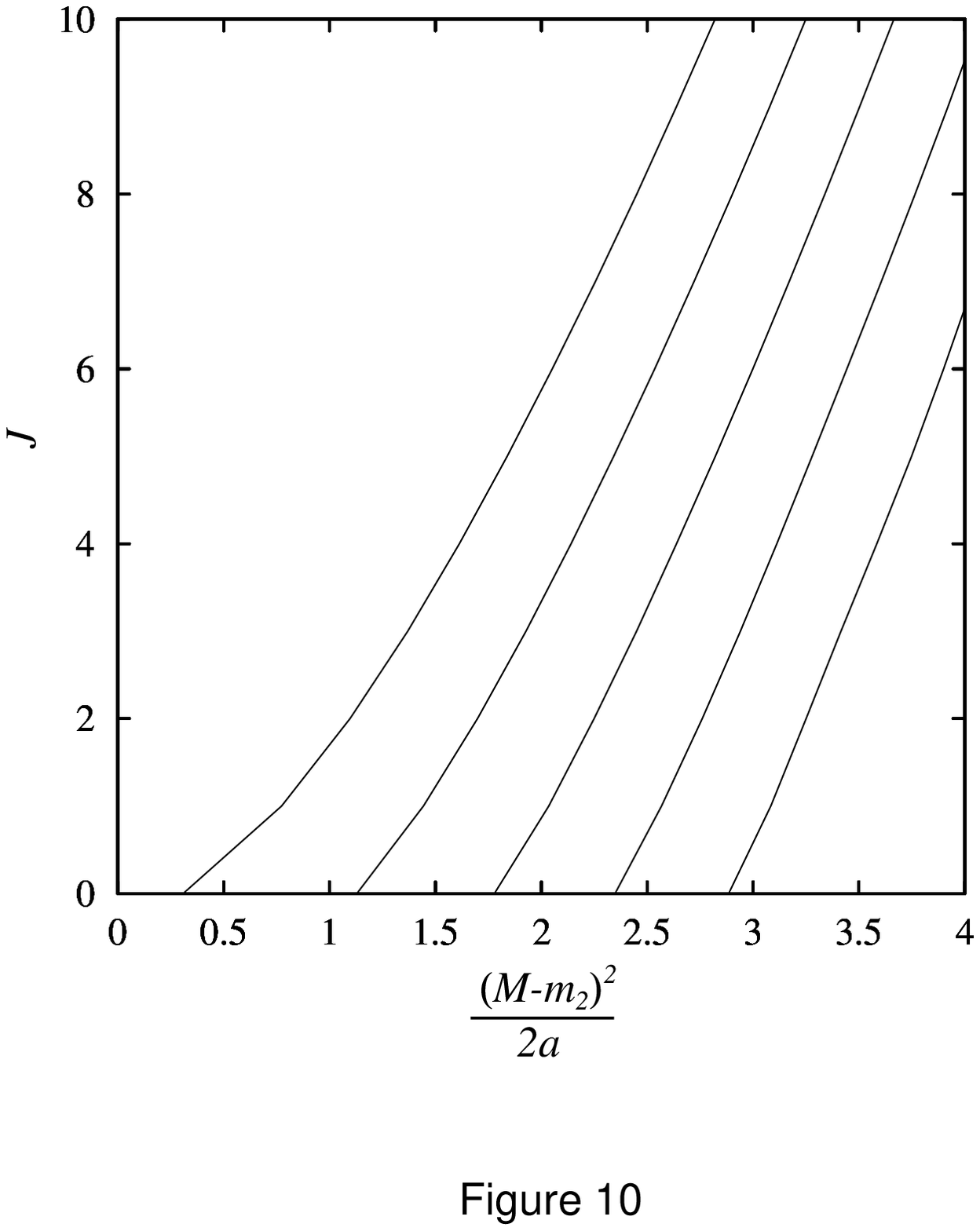}
\end{figure}

\begin{figure}[p]
\epsfxsize = 5.4in \epsfbox{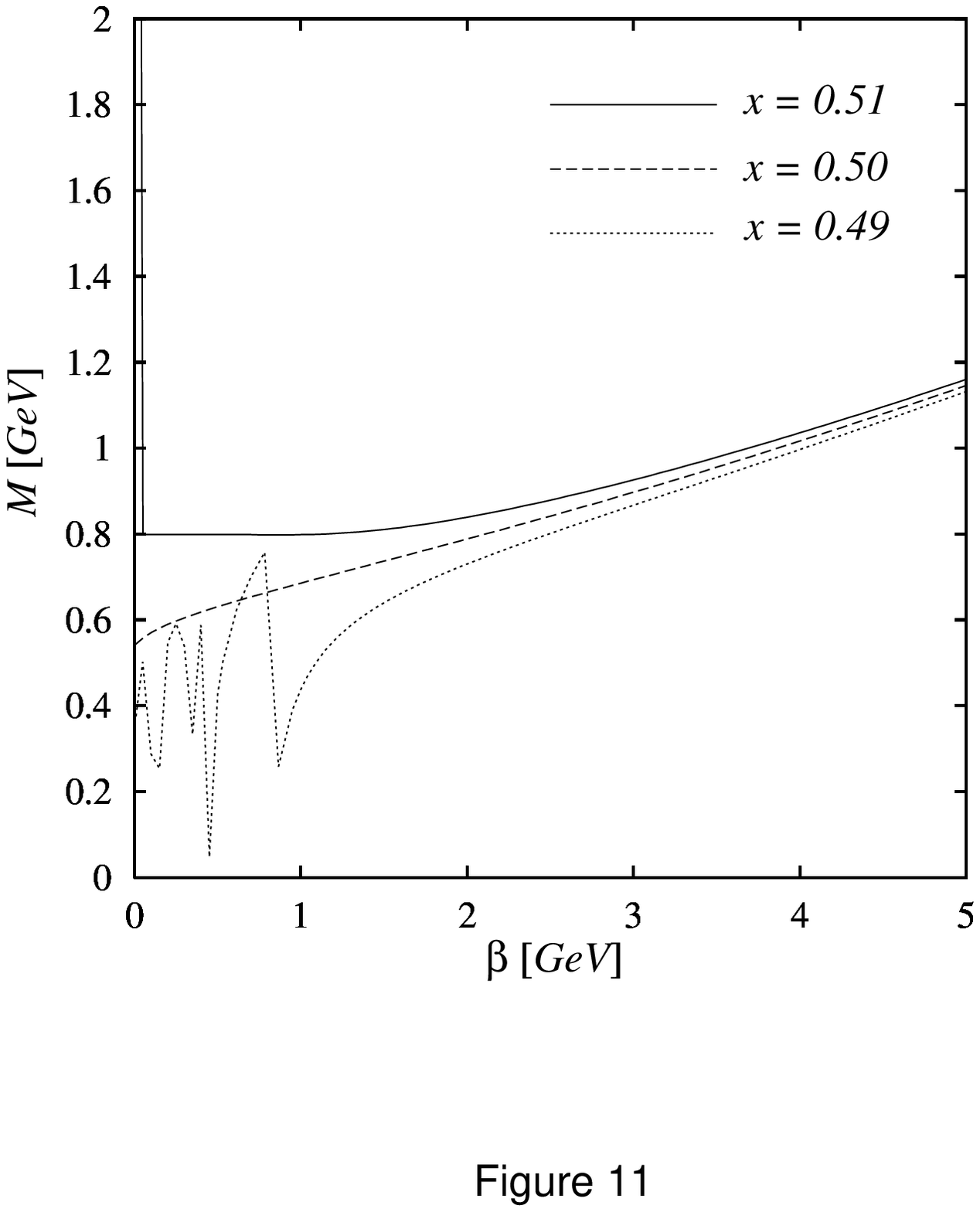}
\end{figure}

\begin{figure}[p]
\epsfxsize = 5.4in \epsfbox{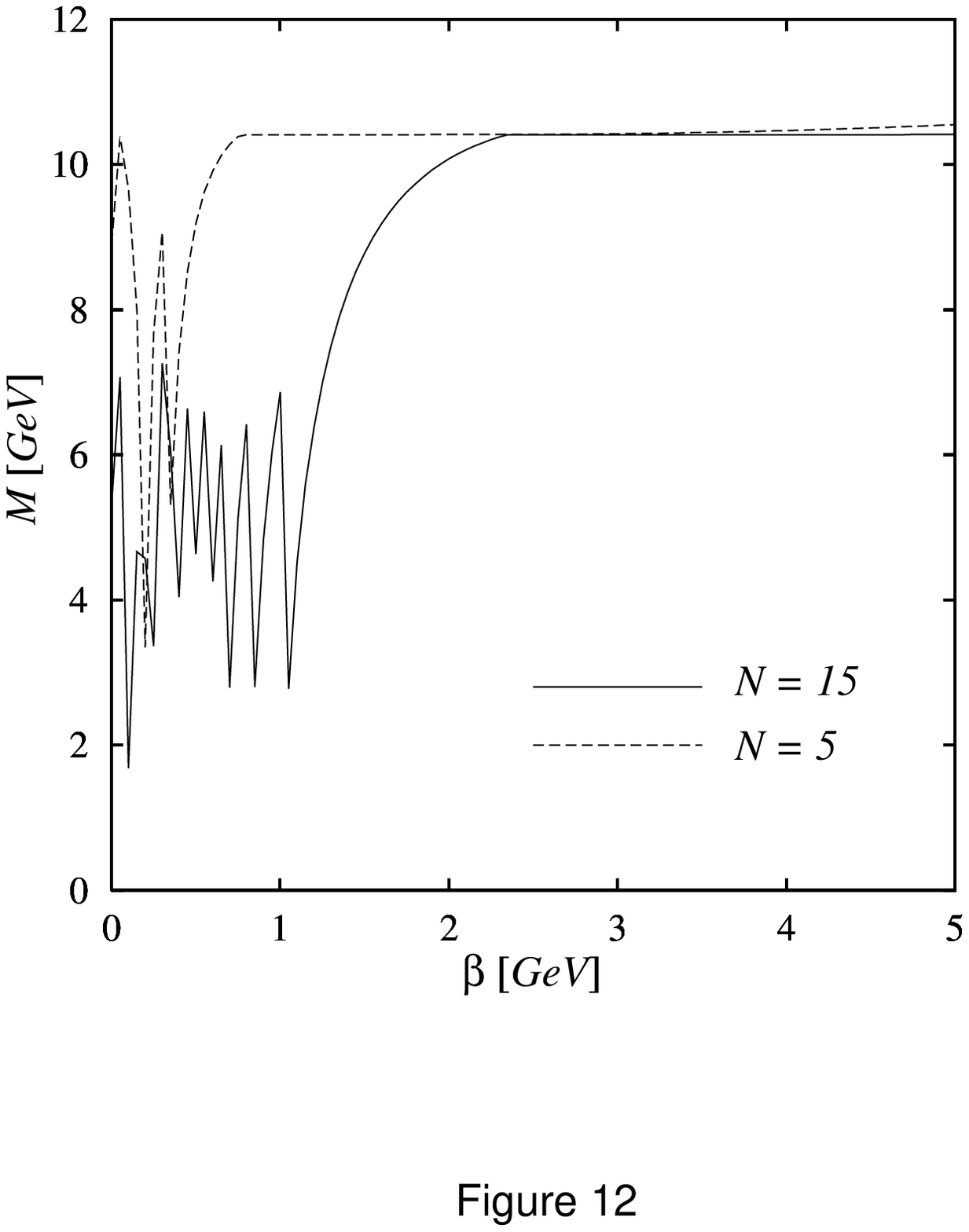}
\end{figure}

\end{document}